\documentclass[aps,prf,preprint,onecolumn,groupedaddress]{revtex4-2}
\usepackage{graphicx}
\usepackage{amsmath}
\usepackage[utf8]{inputenc}
\usepackage[T1]{fontenc}

\usepackage{xcolor,soul}
\usepackage{soulutf8}

\soulregister\citep7
\soulregister\cite7
\soulregister\ref7
\soulregister\eqref7

\begin{document}


\title{Transition from diffusion to advection controlled contaminant adsorption in saturated chemically heterogeneous porous subsurfaces}


\author{Dario Maggiolo}
\email[]{maggiolo@chalmers.se}
\affiliation{Department of Mechanics and Maritime Sciences, Chalmers University of Technology,
Göteborg, SE-412 96, Sweden}

\author{Oskar Modin}
\author{Angela Sasic Kalagasidis}
\affiliation{Department of Architecture and Civil Engineering, Chalmers University of Technology,
Göteborg, SE-412 96, Sweden}


\date{\today}

\begin{abstract}
We show the impact that scalar structures deformation and mixing has on the fate of plumes of waterborne contaminant transported through a chemically heterogeneous, partially adsorbing porous medium, at a typical Péclet number characterizing saturated flows in subsurfaces, $\textit{Pe}= O(1)$. Via pore-scale lattice Boltzmann simulations, we follow the dynamic of a passive scalar injected in a packed bed consisting of a mixture of chemically inert and adsorbing spherical particles. By varying the fraction of the adsorbers $\xi$, randomly and uniformly distributed in the porous volume, we find that the waterborne solute forms concentration plumes emerging between pairs of adsorbing particles. This deformation is a consequence of the different mechanisms of transport characterising the transport of molecules in the proximal and remote pores relative to the adsorbers,  diffusion and advection, respectively. The resulting isoscalar surface embedding the plumes grows at a rate proportional to the average pore-scale velocity $U$ and inversely proportional to the adsorbers' interparticle dimensionless distance, i.e. $\gamma \propto U/\ell_\xi$. We provide a quantification of the characteristic diffusive time scale of the plume $t_\eta \propto \ell_\xi^2/D_m$, which dissipates the concentration differences in the vicinity of the adsorbers, with $D_m$ the molecular diffusion coefficient. Thus, by quantifying the relative importance of the advection-sustained stretching rate $\gamma$ and plume mixing rate $1/t_\eta$ for different values of fraction of adsorbers $\xi$, we establish a transition from diffusion- to advection- dominated macroscopic adsorption, whose time evolutions scale as $\propto \sqrt{t}$ and $\propto t$, respectively. Such a transition is determined by the amount of adsorbers within the medium, with diffusion and advection dominating at high and low fractions, respectively. Our numerical analysis provides $\ell_\xi/d \approx 4/\ln(2) \textit{Pe}^{-1}$ as the critical distance between adsorbers that set the transition, being $d$ the pore size. 
\end{abstract}


\maketitle


\section{Introduction}

Transport of waterborne solutes in porous media occurs in many natural processes and engineering applications. Of particular interest in the context of environmental pollution is the transport of contaminants in subsurfaces, both in agricultural and urban environments. Such pollutants are indeed carried by moving fluids within soil media in agricultural landscapes as well in the urban substrates devoted to the control of rain events, such as green areas and green roofs. In such subsurface environments, a rain event is the usual triggering mechanism, which induces the movement of fluids through the porous spaces, carrying contaminants gathered from the external environment or internally present in the soil media and mobilized by the flow. A notable example of such contaminant sources are fertilizers, which occur as solid particles that dissolve in water infiltrating the subsurface and cause pollution of aquifers and urban water systems. ~\cite{berndtsson2010green,kok2016evaluation}.

A promising solution to prevent such pollution scenarios is the introduction of adsorbers in the soil media, such as biochar, able to adsorb the waterborne pollutants~\cite{ahmad2014biochar,qiu2022biochar}. Biochar not only provides a sink for waterborne contaminants, but also supports plant growth~\cite{mohanty2018plenty}. In this scenario, the porous system becomes strongly heterogeneous in its chemical adsorption capacity, being formed of a mix of chemically inert and adsorbing particles. The identification of an effective strategy for the introduction of adsorbers requires not only knowledge of the specific surface chemistry determining the solute contaminant adsorption onto and into biochar particles, but also of the fluid-dynamic mechanisms that govern the mobilization, transport and spatial redistribution of solute molecules transported within such an heterogeneously adsorbing porous medium.

The transport of a solute in a porous medium is a complex mechanism that may exhibit chaotic features also in laminar flows through three-dimensional homogeneous materials. 
An early-time pore-scale stretching regime dominated by advective forces is usually followed by a coalescence diffusion-induced regime. In the former, scalar elements are deformed and spatially separated to form a filamentary and heterogeneous structure of the transported concentration field. At longer times diffusion brings this structures together, suppressing such an early-time heterogeneous morphology~\cite{villermaux2012mixing,le2015lamellar}. These two regimes are well-separated for high Péclet numbers and the relative importance of advection over diffusion can be enhanced by strong microstructural heterogeneities. For instance, it has been shown how the specific topological traits that distinguish three-dimensional porous materials, a sequence of pore throats in the proximity of grain contact points and enlargements in the cores of pore spaces, trigger stretching and folding of pockets of solute, leading to anomalous dispersion and chaotic advection~\cite{lester2013chaotic,aref2017frontiers}. This anomalous behaviors may persist as long as the advective forces are on the same order of magnitude as molecular diffusion, i.e. for a finite value of the Péclet number $\textit{Pe}\gtrsim5$~\cite{heyman2020stretching}.

If such fluid-dynamic mechanisms responsible of deformation of scalar elements has been shown to be triggered by intense local perturbation of the flow field, such as strong geometrical heterogeneities within the porous medium, less is known about the role of chemical heterogeneities within subsurfaces. Scalar deformation of concentration fields is an important mechanisms because it determines the extent of the diffusive surface within the porous medium, the microscopic concentration gradients and the effective diffusion and adsorption in porous media.
Water infiltration into agricultural soil and urban green substrates during moderately intense rainfall events is of the order of 10 mm/h, which implies that, in the typical millimeter-size pore space of soil media, the water infiltration rate is of the same order of the molecular diffusive rate of contaminants~\cite{liu2019influence}. With the trend of increasing intensity of stormwater events in Northern latitudes~\cite{olsson2018extremregn}, the condition $\textit{Pe}\gtrsim1$ is likely to be encountered even more frequently in such urban and agricultural areas. 
Within this picture, the presence of chemical heterogeneities, such as adsorbing particles, may also act to alter the distribution of concentrations within the pore interstices, possibly deforming the transported scalar elements and affecting the pore-scale microscopic concentration gradients responsible to deliver contaminants to the adsorbing sites via diffusion.

A pore-scale characterization of the transport is thus necessary to understand the behaviour of a waterborne contaminant flowing into a subsurface. Direct visualization of mixing in porous media via high-resolution experimental images at high Péclet numbers has recently confirmed the role of diffusive and advective forces in deforming scalar elements and determining the rate of the chaotic dispersion~\cite{heyman2021scalar}, the mixing of initially separated scalars~\cite{kree2017scalar} and, in turn, the intensity of the chemical gradients that control pore-scale reaction and adsorbing mechanisms~\cite{heyman2020stretching}. Such experiments, complemented with pore-scale numerical studies, have supported the development of mathematical formulations able to predict the scalar structure deformation and mixing processes of transported scalars in homogeneous and heterogeneous media at high Péclet numbers~\cite{le2015lamellar,turuban2019chaotic,souzy2020velocity} and their impact on reactive processes~\cite{ye2020plume}. Nevertheless, research on the impact of pore-scale heterogeneous adsorption on scalar structures deformation and mixing is instead scarce, also at at low Péclet numbers. 

From a macroscopic perspective, adsorption in porous media can be regarded as a filtration process which strongly depends on the pore-scale transport mechanisms, and macroscopic models can be formulated in order to take into account the pore-scale geometrical features that affect local fluid velocities~\cite{miele2019stochastic}. In the presence of chemical heterogeneities of the medium surfaces and of their adsorption processes, predictive macrohomogeneous models for contaminant transport must include the effects of spatial variation of both pore-scale velocities and local adsorption rates~\cite{johnson2003effect}, which may greatly impact the fate of the transported scalar as local disturbances in hydraulic conductivities~\cite{tyukhova2016mechanisms}.
Effective velocity, dispersion, and reaction exhibit a complex dependence on the physical parameters, Péclet and Damköhler numbers, making the prediction of such transport mechanisms challenging from a mathematical point of view~\cite{municchi2020macroscopic}.

 In the present study, we clarify the role that the introduction of chemically adsorbing particles in inert, weakly heterogeneous porous media has on the mechanisms of spatial mixing, diffusion, and macroscopic transport and adsorption of a waterborne contaminant. We will study such a scenario via pore-scale numerical simulations with varying the volume fraction of adsorbents introduced in a homogeneous packed bed for a moderate value of the Péclet number, $\textit{Pe}\sim O(1)$ and typical adsorption rates of waterborne pollutant into carbonaceous material, such as biochar. We will provide a relationship between fraction of adsorbers and deformation processes of the transported scalar, which will allow us to determine the characteristic diffusive and advective time scales for contaminant transport. We will show in which configurations diffusive mixing and avdective transport determines the macroscopic adsorption rate and the contaminant retention in such chemically heterogeneous porous media.

\section{Numerical method~\label{sec:num}}

\subsection{Pore structure generation and surface adsorption}
\begin{figure*}
\includegraphics[width=\textwidth]{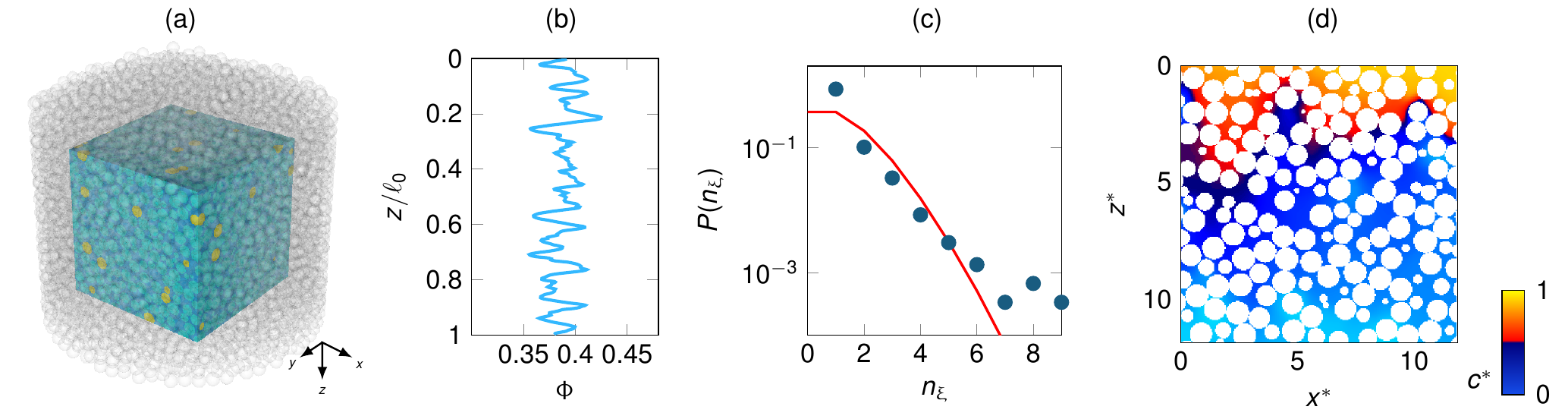}
\caption{(a) Packed bed generation with spherical particles dropped in a cylindrical container (gray particles), within which a cubic domain of size $\ell_0^3=256^3$ computational units is selected (light-blue particles). The parts of particles laying outside the domain are cropped-out and the resulting effective number of particles belonging to the domain is is $n_0=1939$. In each simulation, a fraction $\xi$ of spherical particles is randomly selected as adsorbers (yellow coloured, in the picture for the case $\xi=0.05$). (b) Average porosity $\Phi$ at each $(x,y)$ plane, which shows the homogeneity of the domain along the streamwise direction $z$. (c) Probability distribution function of number of adsorbing particles $n_\xi$ contained within randomly chosen spherical boxes of diameter $\ell_\xi$ (blue marks). The red line depicts the Poisson distribution for an average occurrence of events $\langle n_\xi \rangle = 1$. The observed discrepancies between the measured probability and the Poisson distribution $n_\xi$ can be ascribed to finite-size effects, induced by the symmetric boundary conditions along the transverse directions, and by the finite particle size that impedes overlapping. (d) Snapshot of solute transport and adsorption, visualized via dimensionless concentration $c^*$ injected from the top plane at $z^*=0$ with inlet concentration $c_0^*=1$. Note the concentration sheet $\Sigma$ at $c^*=1/2$ visible from the sharp change in color of the solute. \label{fig1}}
\end{figure*}

We generate a packed bed sample by solving the rigid body dynamics of falling spherical particles within a cylindrical container~\cite{boccardo2015validation,pettersson2020impact}. A cubic domain is then selected within the container and the volumetric space is discretized in equally sized voxels, so that the spherical particle diameter and the domain size are $d=21.6$ and $\ell_0^3 = 256^3$ computational nodes, respectively. We check planar porosity values along the packing direction $z$ to ensure that the selected volume is sufficiently distant from the container walls and it is homogeneous, see Fig.~\ref{fig1}, panels (a) and (b). The volumetric porosity is $\Phi=0.39$, from which we calculate the effective number of particles contained in the volume space as $n_0={\ell_0^*}^3\ (1-\Phi)6/\pi=1939$, with $\ell_0^*=\ell_0/d$ the dimensionless domain size along the three Cartesian directions.
We compute the pore size distribution via a watershed algorithm and extract an equivalent average pore diameter $\approx d$. 
For each simulation case, we select a random fraction $\xi$ of spherical particles (via a random uniform permutation of particles indexes) and along their surface $S_\xi$ we assign a solute adsorbing rate expressed via the following, partially adsorbing, first-order reaction kinetics:
\begin{equation}
\textit{Da} \ c^* \big |_{S_\xi} = -\frac{\partial c^*}{\partial \lambda_s^*} \bigg |_{S_\xi} \ , \label{eqads}
\end{equation}
where $\textit{Da}=k d^2/D_m$ is the Damköhler number characterizing the ratio between adsorption rate $k$ and diffusion rate $D_m/d^2$ (with  $D_m$ the molecular diffusion), $c^*=c(\boldsymbol{x},t)/c_0$ is the dimensionless concentration at position $\mathbf{x}=(x,y,z)$ and time $t$ that refers to the  injected concentration $c_0$, and $\lambda_s^*=\lambda_s/d$ is the dimensionless direction pointing inward to the adsorbing particle surface. The adsorbing spherical particles are randomly and uniformly distributed in the volume space with an average dimensionless interparticle distance $\ell_\xi^*=\ell_\xi/d=\ell_0^*/\sqrt[3]{\xi n_0}$. Thus, the probability of having a certain number of adsorbing particles $n_\xi$ within an arbitrary spherical domain of radius $\ell_\xi/2$ follows a Poisson distribution with average rate  $\langle n_\xi \rangle =1$, as confirmed by a spherical-box counting computation performed in the porous domains, shown in Fig.~\ref{fig1} (c).

\subsection{Numerical simulations of pore-scale transport}

We investigate solute transport and adsorption into media with variable adsorption capacity, in order to mimic the process of waterborne contaminant treatment by means of the introduction of adsorbers into the subsurface. We consider 11 simulation cases, 6 simulations cases with an increasing fraction of adsorbers $\xi=1/160$, $1/80$, $1/40$, $1/20$, $1/10$, $1/5$ and $\textit{Da}=1.3$ and 5 simulation cases with $\xi=1/160$, $1/80$, $1/40$, $1/20$, $1/10$ and $\textit{Da}=2.6$. 
The values of $\textit{Da}= O(1)$ are chosen to represent the balance between adsorption rates typically observed in biochar batch adsorption experiments, of the order of 1 mg/L per 100 mg/L of solution over a minute, i.e. $k= O(10^{-3}\ \mathrm{s}^{-1})$, and the diffusion rate of species in liquids within a typical 1mm pore space, i.e. $D_m/d^2 = O(10^{-3}\ \mathrm{s}^{-1})$.
For each simulation case, 3 random arrangements of the adsorbing particles are considered, for a total of 33 simulations. The computed data that follows in the manuscript thus refer to the ensemble averaged values among such different realizations.  The advection-reaction-dispersion equations (ARDE) for the transport of the scalar concentration $c^*(\boldsymbol{x},t)$ is solved for a steady-state flow field, whose dimensionless solenoidal $j$-th Eulerian velocity component is $u_j^*=u_j(\boldsymbol{x})/U$. The ARDE reads as: 
\begin{equation}
\frac{ \partial c^*(\boldsymbol{x},t)}{\partial t^*} +   \frac{ \partial c^*(\boldsymbol{x},t) u^*_j(\boldsymbol{x})}{\partial x_j^*} = \frac{\partial }{\partial x_j^*} \bigg (  \frac{1}{Pe} \frac{\partial c^*(\boldsymbol{x},t)}{\partial x_j^*} \bigg ) \ , \label{eqade}
\end{equation}
with $U$ the intrinsic average flow velocity in the medium, $x^*_j=x_j/d$ the dimensionless $j$-th direction, $t_a=t/t^*=d/U$ the pore-scale characteristic advective time and $\textit{Pe}=Ud/D_m$ the Péclet number.
Equation~\eqref{eqade} is solved via the lattice-Bolztmann methodology~\cite{succi2001lattice}.
The steady-state flow solution is achieved by solving a first lattice population within the void space of the porous medium, with no-slip boundary conditions at the fluid-solid interface and a pressure gradient acting to force the flow along $z$. At the boundaries along the transverse directions $x$ and $y$, the symmetry is guaranteed with free-slip boundary conditions, while we impose periodic conditions along the streamwise direction $z$. A second population is then computed for solving the transport of the passive scalar $c^*$ with constant injection $c^*(z=0)=1$ and its adsorption at the boundaries $S_\xi$, according to Eq.~\ref{eqads}~\cite{maggiolo2020solute}. At the rest of the fluid-solid boundaries and at the outlet a zero-gradient condition is imposed. The applied pressure gradient that drives the flow is chosen to achieve a Péclet number representative of the balance between the infiltration rate into soil media during an intense rain event and the diffusion rate of solute within the pore space. The intrinsic average flow velocity in a soil medium during an intense event is of the order of 10 mm/h~\cite{liu2019influence}. The infiltration rate in a typical pore space $d\sim 1\ \mathrm{mm}$ thus results $U/d = O(10^{-3} \ \mathrm{s}^{-1})$, on the same order of the diffusive rate, and the Péclet number 
$\textit{Pe}=1.72 = O(1)$. The resulting  behaviour of the constantly injected solute concentration will depend on the transport and adsorption mechanisms occurring within the pore space, as illustrated  and intuitively inferable from Fig.~\ref{fig1} (d).
Further numerical details and a validation of the computational methodology are provided in the Appendix~\ref{app:lbm}.

\section{Results}

 \begin{figure}
\noindent\includegraphics[width=0.95\textwidth]{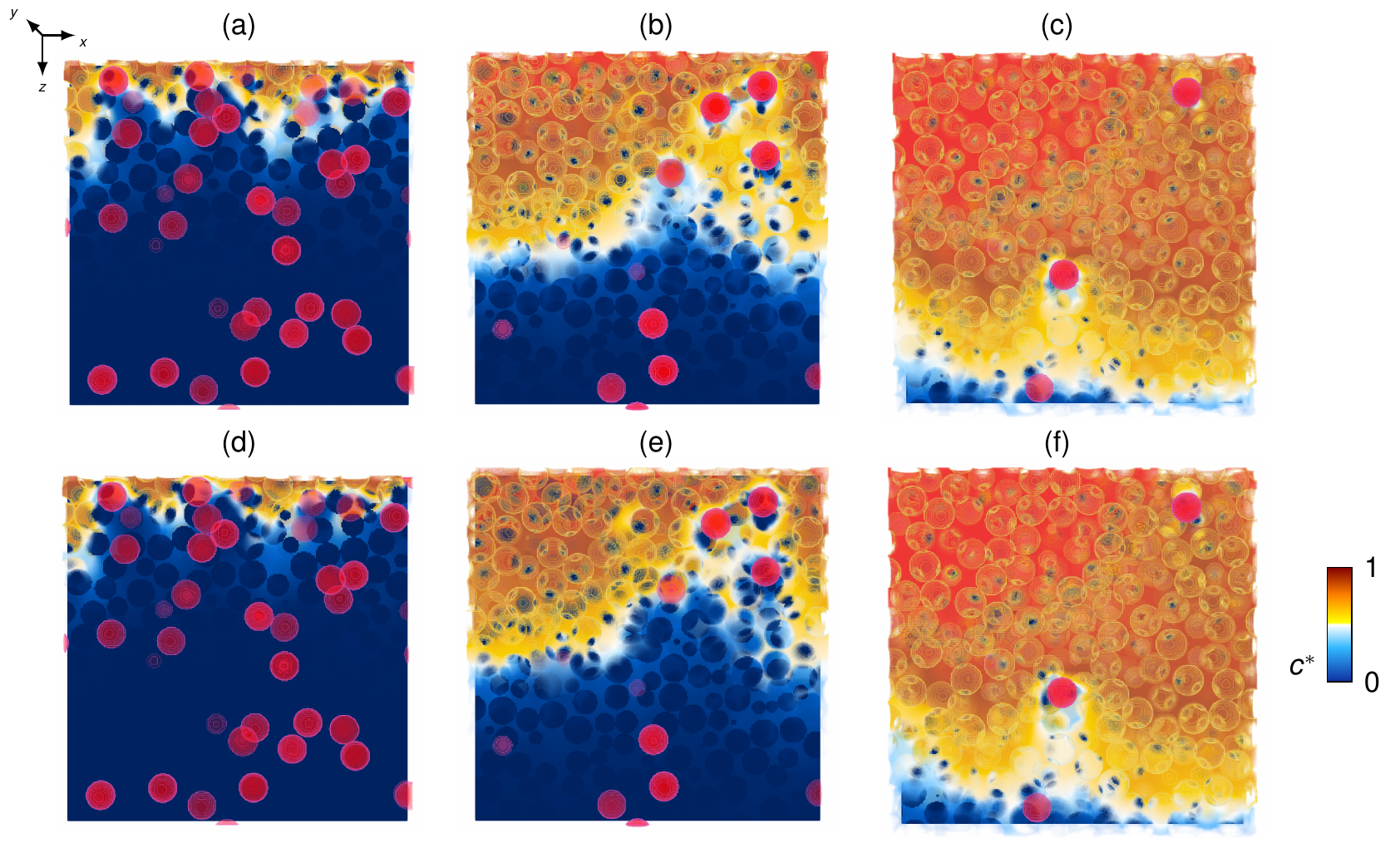}
\caption{\label{fig2} Snapshots of spatial distribution of concentration $c^*$ injected from the top boundary $z^*=0$ at a characteristic surface growth time $t=2\ell_\xi/U$. The snapshots are taken at $y=\ell_0/2$ and reported for (a) $\textit{Da}=1.3$ and $\ell_\xi^*=0.1$, 
(b) $\textit{Da}=1.3$ and $\ell_\xi^*=0.025$, (c) $\textit{Da}=1.3$ and $\ell_\xi^*=0.00625$, (d) $\textit{Da}=2.6$ and $\ell_\xi^*=0.1$, (e) $\textit{Da}=2.6$ and $\ell_\xi^*=0.025$, (f) $\textit{Da}=2.6$ and $\ell_\xi^*=0.00625$. The adsorbing particles contained within the porous volume defined within $y=[\ell_0/2, \ell_0/2+\ell_\xi/2]$ are represented with red-colour circles. Plumes of solute are shaped by the presence of adsorbing particles, emerging across $(x,y)$ cross-sectional fluid areas, defined between pairs of adsorbing particles, distanced, on average, $\ell_\xi$. At a time $t\propto \ell_\xi/U$, plumes surface growth follows the extrusion of the $n_2\propto \ell_\xi^{-2}$ cross-sectional fluid areas of perimeter $\pi_\xi \propto \ell_\xi$, the streamwise elongation scales as $h_\xi \propto \ell_\xi$ (as qualitatively visible from the above figure), and the total surface growth results statistically constant as $\Sigma_\xi^*\ell_0^*-1 \propto h_\xi \pi_\xi n_2$, see also Eq.~\ref{eqstreprop}.}
\end{figure}

\subsection{Plume stretching and surface growth}

\begin{figure}
\noindent\includegraphics[width=0.95\textwidth]{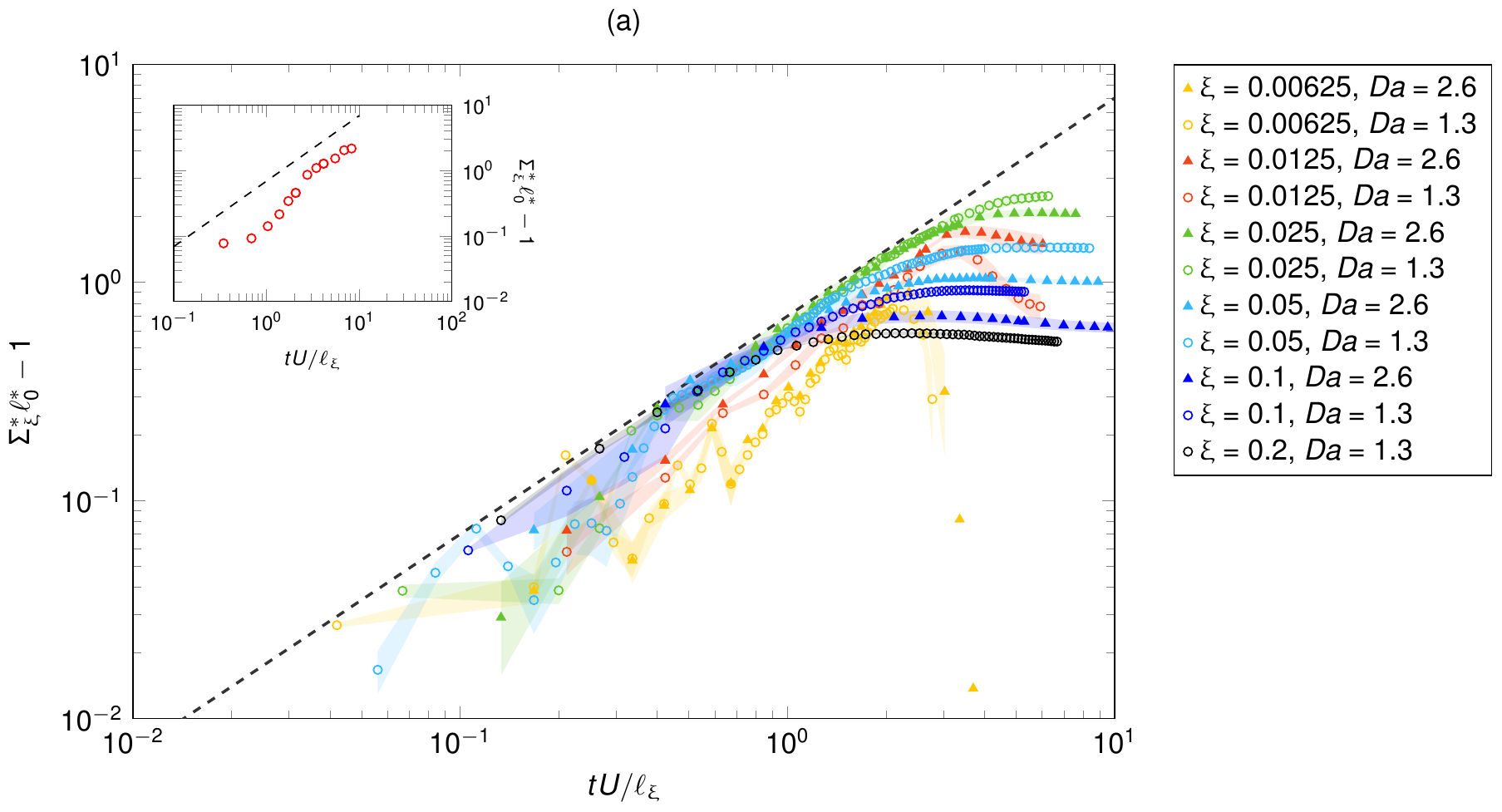}
\caption{(a) Evolution of the dimensionless concentration sheet per fluid volume $\Sigma^*_\xi$, computed at $c^*=1/2$, at dimensionless times $t U /\ell_\xi$, with $\ell_\xi/U$ indicating the characteristic advecting time for bulk transport between adsorbers. Circular and triangular marks address data for $\textit{Da}=1.3$ and $2.6$, respectively, while different colours denote different adsorbing particle concentration $\xi$. The shaded areas indicate the standard error computed between the 3 different random realizations for each case. A linear time-dependent surface growth is observed, following $\Sigma^*_\xi\ell_0^*-1 = \gamma t$, with $\gamma=\gamma^* U/\ell_\xi$ the surface growth rate, $\gamma^*$ a constant value, and  $\ell_0^*$ the dimensionless length of the domain. The black dashed lines indicates the solution of Eq.~\eqref{eqstretch}, for $\gamma^*=\ln(2)$. 
Inset: the linear time-dependent trend is confirmed at a small value of $\xi=0.00625$ in an additional simulation performed in a larger computational domain $2\ell_0^3$ (red marks).}
\label{fig3}
\end{figure}

We follow the temporal evolution of the concentration of the transported scalar $c^*$. The scalar concentration injected from the top boundary $z^*=0$ is transported and mixed within the pore interstices, forming plumes of solute concentration whose characteristic size is affected by the distribution of the adsorbing particles, placed, on average, at a relative distance $\ell_\xi^*$ within the porous medium. Concentration plumes emerge across orthogonal fluid spaces whose perimeters are defined by transversally aligned pairs of adsorbing particles, as for instance depicted in Fig.~\ref{fig2}. We found that such plumes experience a a stertching process developing linearly in time, which is induced by the difference between bulk mass transport in the vicinity of the adsorbing particles, which is hindered by adsorption, and the transport in the pores farther from such adsorbing sites, where the concentration is freely advected. 
We report the growth of the plumes surface $\Sigma$, i.e., its stretching, by computing the extent of the scalar sheet to which corresponds a concentration value $c^*=1/2$, i.e. an isoscalar surface. Thus $\Sigma$ represents the backbone concentration sheet at half-way between maximum concentration and null-concentration areas and with $\Sigma^*_\xi= (\Sigma/d^2) /(\Phi{\ell_0^*}^3)$ we indicate the dimensionless specific surface area of such a sheet per volume of void space. At the initial time $t=0$, the solute surface corresponds to a sharp front placed at the inlet cross section $\Phi{\ell_0^*}^2$, thus leading to $\Sigma^*_\xi(t=0)={\ell_0^*}^{-1}$.
In Fig.~\ref{fig3} (a) we report the computation of $\Sigma^*_\xi$ showing the evolution of the solute plume surface as a consequence of a linear time-dependent stretching process:
\begin{equation}
\Sigma^*_\xi (t)= \dfrac{1}{\ell_0^*} (1+\gamma t  )  \ , \label{eqstretch}
\end{equation}
where $\gamma$ is the stretching rate:
\begin{equation}
\gamma=\gamma^*\dfrac{U}{\ell_\xi} \ , \label{eqrate}
\end{equation} 
and $\gamma^*$ is a constant prefactor, which, via a linear fitting, we determine close to $\gamma^*\approx\ln (2)$. The stretching rate $\gamma$ results proportional to the average pore-scale mass transport rate $t_a^{-1}=U/d$ and inversely proportional to the characteristic length $\ell_\xi^*$, i.e. $\gamma \propto (t_a \ell^*_\xi)^{-1}$. Thus, the plume surface growth per fluid volume occurs faster for adsorbing particles placed at shorter distances. This proportionality is a direct consequence of the linear growth process. Solute plumes emerge between pairs of adsorbing particles placed, on average, along a perimeter proportional to their average distance, i.e. $\pi_\xi \sim \Phi \ell_\xi^*$. The concentration sheet $\Sigma$ grows following the extrusion of such a perimeter line advected at a rate $t_a^{-1}$ along the streamwise direction. The average number of evolving plumes in the two-dimensional transverse plane $(x,y)$ is $n_{2}=(\ell_0^*/\ell_\xi^*)^2$
 and the concentration sheet growth rate per unit inlet fluid area can be derived as:
 \begin{equation}
 \gamma \propto t_a^{-1} n_2\pi_\xi / {\Phi \ell_0^*}^2 =  (t_a \ell^*_\xi)^{-1}  \ . \label{eqstreprop}
 \end{equation}

We can interpret Eq.~\eqref{eqstreprop} as the measure of a kinetic roughening process of a scalar element, whose extent along the longitudinal direction--its roughness height--is induced by the flow velocity $\propto t_a^{-1}$ and whose transversal periodicity--its lateral correlation length--is proportional to the average adsorbers distance $\ell_\xi^*$. This surface growth behaviour resembles a kinetic roughening process statistically constant over a length $\ell_\xi^*$.
 We also note  in Fig.~\ref{fig3} (a) that for low values of fraction of adsorbers, i.e. $\xi<0.025$, the computed values of $\Sigma$ exhibit a sharp decay at long times, because the plume sheet $\Sigma$ escapes the porous volume. However, we note that the linear trend holds for all cases, at least as long as the surface $\Sigma$ is contained within such a volume.
 
 We also highlight that $\gamma^*U$ is the average solute velocity for a plume of concentration $c^*=1/2$. At a time $t$, such a plume had travelled a distance along $z$ that equals:
 \begin{equation}
 h_\xi^*(t)\equiv \gamma^* U t/d = (\Sigma_\xi^*(t)\ell_0^*-1) \ell_\xi^* \ , \label{eqdepth}
 \end{equation}
 where the right hand side is recovered from Eq.\eqref{eqstretch}.

\begin{figure}
\noindent\includegraphics[width=0.99\textwidth]{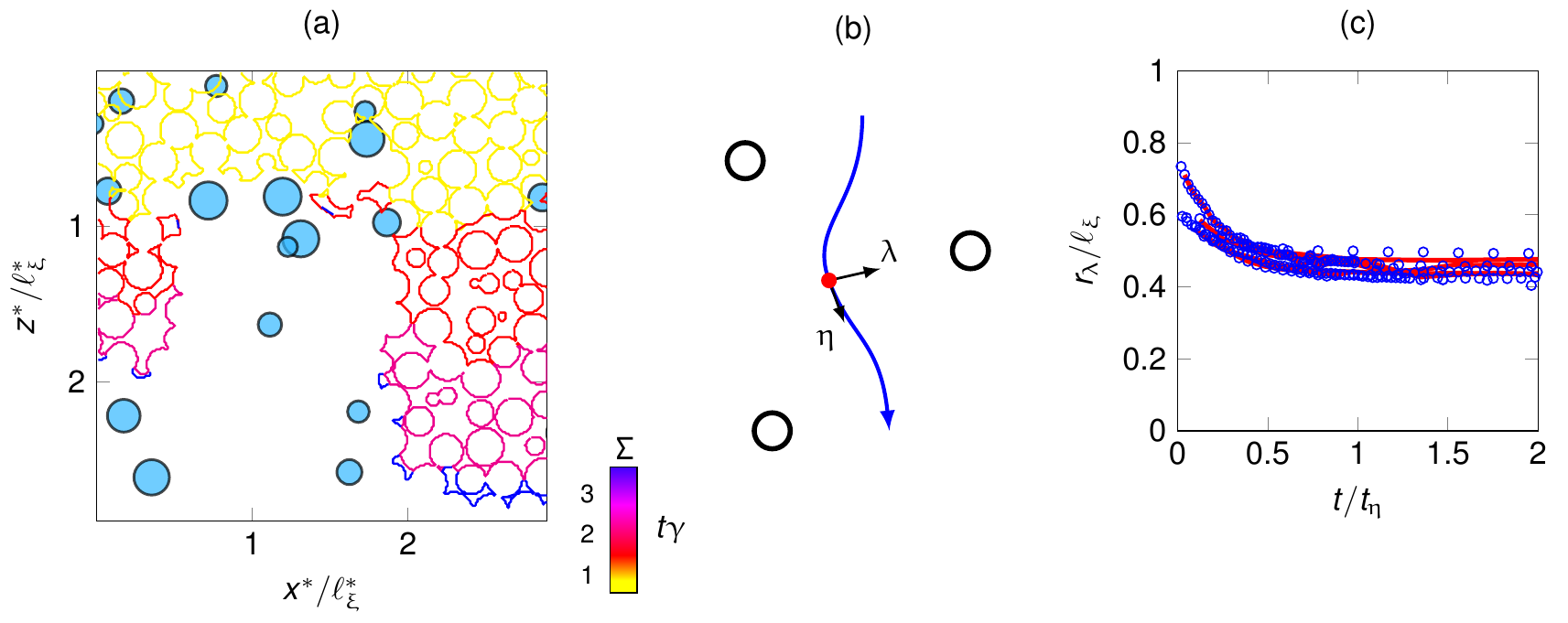}
\caption{(a) Visualisation of plume sheet surface evolution at different dimensionless times $\gamma t$ (colors) for a cross section $(z,x)$ of a case with fraction of adsorbers $\xi=0.05$. The linear time-dependent surface growth process extrudes a plume along its perimeter proportionally to $\ell_\xi$ and $1/\gamma$. Particle adsorbers within a distance to the cross-section $\pm 10d$ are sketched in light blue, and their radius is made inversely proportional to such a distance, to highlight their impact on solute transport.  (b) Sketch showing the evolution of a scalar point at the isoscalar surface $\Sigma$, under the concurring effects of advection and diffusion, with indicated the coordinates $\eta$ and $\lambda$. (c) Average shortest distance between $\Sigma$ and adsorbers for $\textit{Da}=1.3$ (blue markers) and $\textit{Da}=2.6$ (red lines), which results $r_\lambda = \ell_\xi/2$.}
\label{fig4}
\end{figure}

Given the low value of the Péclet number, the typical pore-scale advective stretching rate $d/U$ is rapidly suppressed by diffusion that brings together scalar elements.
The observed linear time-dependent stretching and surface growth of solute plumes suggests that concentration plumes onset is firstly determined by diffusion, which homogenized the pore-scale transport, mixing the solute at the pore-scale, while it later follows a linear deformation induced by the presence of adsorbers, which differentiates the mass transport behaviours in their proximal and remote pores.
Within this picture, we are thus allowed to represent the solute evolution as a statistical average of independent, well-mixed, individual  plumes, which weakly interact with each other and deforms under the action of spatially varying transport rates within the fluid volume available between adsorbing particles. The evolving shape of such individual plumes is well captured by snapshots of our simulations depicted in Figs.~\ref{fig2} and \ref{fig4}(a).  It takes approximately a time $1/\gamma$ (i.e. the average time that takes the solute to encounter an adsorbing particle) to observe such a plume formation mechanism, after which the solute concentration is transported along spaces defined by random pairs of adsorbing particles, with clear effects of voiding and clustering of such particles on the solute plume evolution.

\subsection{Diffusion broadening at short and intermediate times}

As we have discussed in the previous section, given the low value of the Péclet number and the weak geometrical heterogeneity that characterizes our porous medium (being composed of equally sized spherical particles), we expect to rapidly observe a coalescence regime, where individual pore-scale scalar structures are homogenized by diffusion~\citep{villermaux2012mixing}. In this conditions, it is the presence of spatially varying adsorption rates that yield the observed surface growth mechanism. The latter arises from the difference between scalar transport along different pores: within the pores proximal to the adsorbers, solute transport is governed by diffusion and adsorption directed along the normal to the adsorbers surface, whereas within more remote pores bulk concentrations are transported by advection. This mechanism triggers the observed solute plume growth and stretching along the space available between pair of adsorbers.

 Advection within the remote pores primarily acts along a direction tangential to the adsorbing surface. At the same time, diffusion brings solute species from the plume core to the adsorbers surface, along a  normal direction. We refer these two directions with $\eta$ and $\lambda$, respectively, see the sketch in Fig.~\ref{fig4}(b).
For an instantaneous step concentration input, along a generic direction $x_j$, diffusion broadening can be described by Fick's law solution:
\begin{equation}
c^*(x_j,t) = \dfrac{1}{2} - \dfrac{1}{2} \mathrm{erf}\bigg(  \dfrac{x_j-Ut}{\sqrt{4 D_mt}} \bigg ) \ ,
\end{equation}
and the absolute value of the dimensionless gradient at $x_j^*=Ut/d$ follows as:
\begin{equation}
\dfrac{\partial c^*(x_j,t)}{\partial x_j^*}  =  \dfrac{1}{2\sqrt{\pi} \sigma^*}  \ , \label{eqdiff0}
\end{equation}
where we make use of the diffusion length
$\sigma^* = \sqrt{D_m t}/d$.

\begin{figure}
\noindent\includegraphics[width=\textwidth]{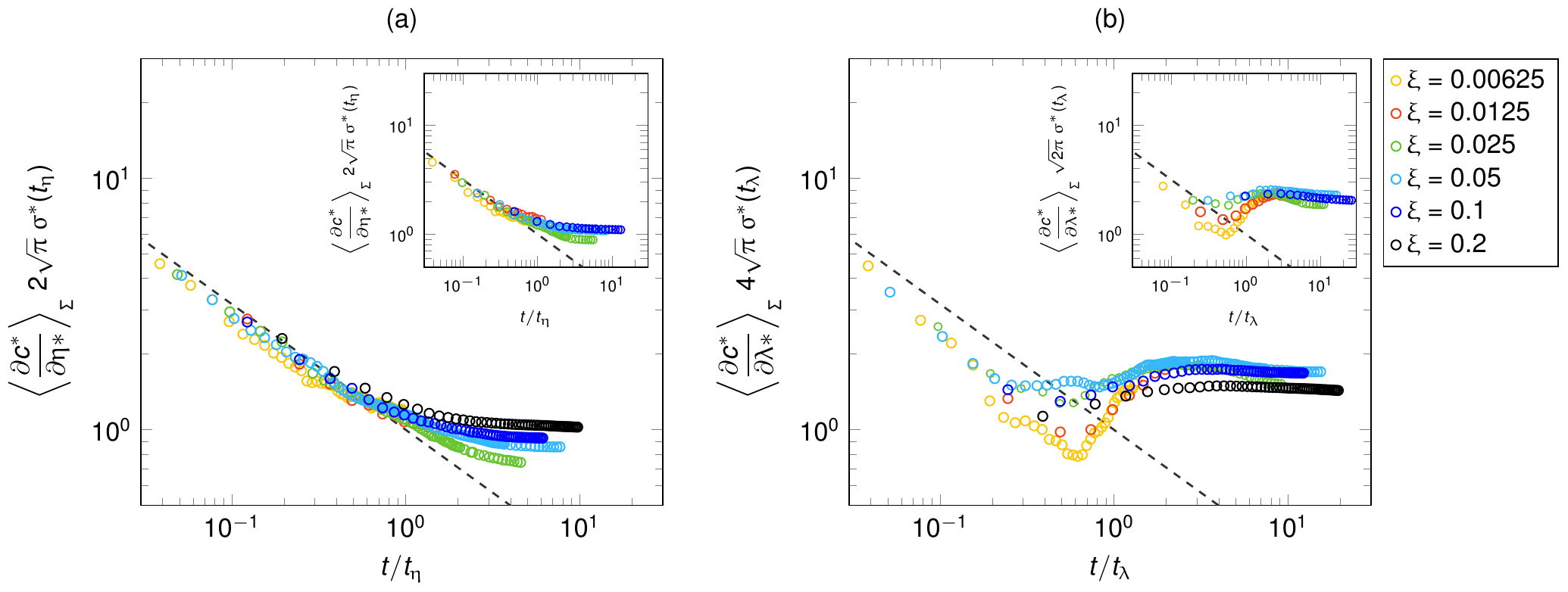}
\caption{Evolution of the average solute concentration gradient computed along the plume surface $\Sigma$. (a) Gradient component along the tangential direction $\eta$, $\langle \partial c^*/\partial \eta^*\rangle_\Sigma$, exhibiting equilibrium at $t_\eta=\ell_\xi^2/2D_m$, for both $\textit{Da}=1.3$ and $\textit{Da}=2.6$ (inset). (b) Gradient component along the normal direction $\lambda$, $\langle \partial c^*/\partial \lambda^*\rangle_\Sigma$, exhibiting equilibrium at $t_\lambda=\ell_\xi^2/8D_m$, for both $\textit{Da}=1.3$ and $\textit{Da}=2.6$ (inset). The dashed lines indicate
 the solution of Eq.~\eqref{eqdiff0} with a prefactor $1/2$ in the case of normal direction $\lambda$ (right panels).
}
\label{fig5}
\end{figure}

We thus compute the average gradient on the concentration sheet $\Sigma$ (indicated by the averaging operator $\langle \cdot \rangle_\Sigma$), along the tangential and normal directions, via the projection of the local gradient along the segment connecting the local position with the closest adsorbing particle centres. 
In Fig.~\ref{fig5} we show the result of such a computation. In particular,
in the insets of Fig.~\ref{fig5}, we show that Eq.~\ref{eqdiff0} very well describe diffusion broadening along $\eta$, i.e. the direction tangential to the adsorbers. Along the direction normal to the closest adsorber, i.e. $\lambda$, we observe a similar behaviour provided that Eq.~\ref{eqdiff0} is multiplied by a prefactor $\sim 1/2$. Such a prefactor reminds that diffusion acts to broaden the plume along two symmetric directions, $\pm \lambda$.

{\hl We also observe that an equilibrium state is found after the initial diffusion broadening, where $\sigma^*$ and the concentration gradients tend to a constant value. 
We relate such an equilibrium to the presence of adsorbers. In particular, let us consider a contaminant molecule travelling within the medium. As a consequence of the continuous plume stretching and bulk transport mechanism, the molecule travels between the adsorbers, at an average distance $\ell_\xi/2$ to the closest adsorber, see the sketch in Fig.~\ref{fig4}(b). The computation of the distance between the isoscalar $\Sigma(c^*=1/2)$ to the closest adsorbers confirms such a value of the average distance, as reported in Fig.~\ref{fig4}(c). Consequently, diffusion broadening mixes the plume and dissipates the concentration differences within a volume embedded between adsorbers, whose radius is $\sim \ell_\xi/2$ and which is advected along the tangential direction $\eta$.

For an instantaneous step concentration input, the typical diffusive time that dissipates the concentration differences within such a moving volume can be found via the unidirectional diffusion equation along the positive direction $\eta$, $\sqrt{D_mt_\eta} = \ell_\xi/2$, which gives:}
\begin{equation}
t_\eta = \dfrac{\ell_\xi^2}{4D_m} \ . \label{eqtime1}
\end{equation}
Along the other direction, since diffusion occurs symmetrically along $\pm \lambda$, the characteristic time results from $\sqrt{2 D_mt_\eta} = \ell_\xi/2$, that is:
\begin{equation}
t_\lambda = \dfrac{\ell_\xi^2}{8D_m} \ . \label{eqtime2}
\end{equation} 
Figure~\ref{fig5} shows that these two characteristic times very well determine the onsets of an equilibrium state in the plume concentration gradients.
This is an interesting result because it provides a relationship between chemical heterogeneity length scale and adsorption mechanisms in the vicinity of the concentration plume.
At times $t<t_\lambda$ adsorption at the adsorbers surface is primarily diffusion-controlled, being limited by the amount of species carried by diffusion. At longer times, in particular $t>t_\eta$, adsorbtion becomes reaction-controlled and a balance is found between diffusion and adsorption, which stops diffusion broadening in the vicinity of the plume.

We also noticed that for the considered values of $\textit{Da}=O(1)$, the particle adsorption rate has a minor effect on the concentration gradients and plume mixing mechanism.

\subsection{Transition from diffusion- to advection- controlled macroscopic adsorption~\label{sec:mod}}

\begin{figure}
\noindent\includegraphics[width=0.79\textwidth]{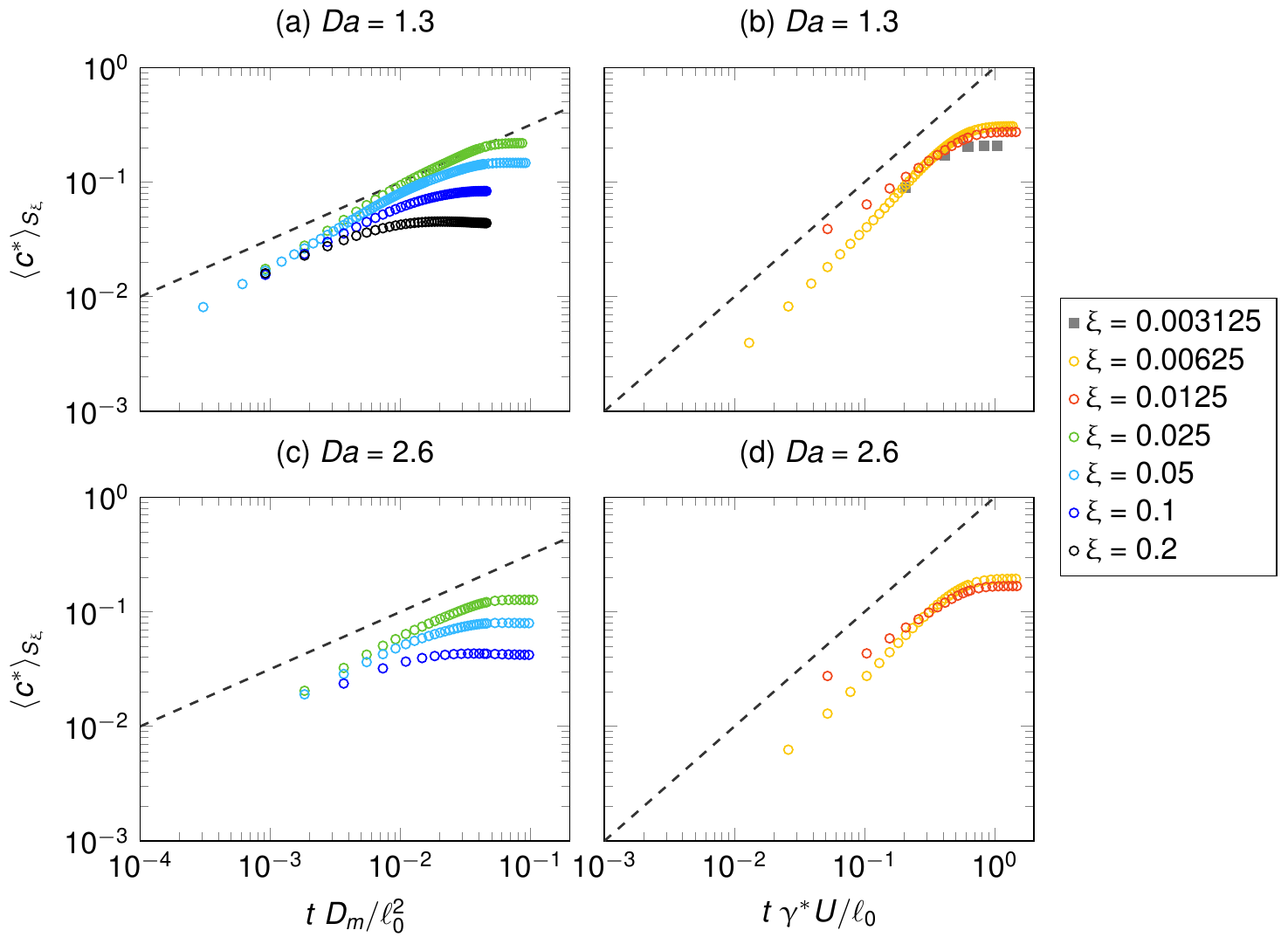}
\caption{
Evolution of the averaged concentration along the surface of the adsorbers $\rangle c^* \rangle_{S_\xi}$. We observe a diffusion-dominated adsorption $\propto \sqrt{t}$ at high fraction of adsorbers $\xi$ (left panels, with $\textit{Da}=1.3$ on the top and $\textit{Da}=2.6$ on the bottom) and an advection-dominated adsorption for low fractions (right panels). The macroscopic characteristic diffusive and advective time are used to scale the x-axis, while the dashed lines represents the solutions provided in Eqs.~\eqref{eqadsmacro1} and \eqref{eqadsmacro2}. An additional simulation performed for $\xi=0.003125$ (gray squares) confirms the linear trend for advection-dominated configurations. The transition between the two adsorption functioning is well predicted by Eq.~\eqref{eqtrans}, which set the critical fraction for the transition at $\xi \approx  0.02$. 
}
\label{fig6}
\end{figure}

At the macroscopic scale, the onset of diffusion broadening equilibrium results in a change of adsorption regime in the proximity of the plume. To complete the picture and take into account the adsorption occurring along all the available surface provided by the adsorbers, i.e. $S_\xi=\xi n_p \pi d^2$, one should also consider the bulk transport mechanism induced by advection, which is quantified via the plume stretching or advective rate $\gamma$. If such a rate is slower then the characteristic plume diffusive rate, the macroscopic adsorption is diffusion-dominated and the transport and mixing of species is governed by characteristic diffusive rates. In such a situation, we may intuitively think that the rate of adsorption is roughly proportional to the time taken by diffusion to bring molecules to the adsorbing sites. The macroscopic characteristic diffusive time scale is $\ell_0^2/D_m$, being $\ell_0$ the characteristic size of the macroscopic volume, and the average amount of mass brought to the global adsorbing surface should scale as:
\begin{equation}
\langle c^* \rangle_{S_\xi} \propto \sqrt{\dfrac{D_m}{\ell_0^2} \ t} \ , \label{eqadsmacro1}
\end{equation}
with $\langle \cdot \rangle_{S_\xi}$ indicates the averaging operator over the available adsorbing surface ${S_\xi}$. On the other hand, when the advective rate is faster than plume diffusive mixing we may write an advection-dominate law for the macroscopic adsorption. In this situation, the bulk transport of the plume with characteristic time $\ell_0 /(\gamma^* U)$ governs the transport of molecules to the adsorbing sites and we infer:
\begin{equation}
\langle c^* \rangle_{S_\xi}\propto \dfrac{\gamma^* U}{\ell_0}\ t \ . \label{eqadsmacro2}
\end{equation}
Because the plume diffusive mixing is determined by the slowest diffusive rate, we may expect Eq.~\eqref{eqadsmacro1} to hold when $\gamma< 1/t_\eta$, whereas Eq.~\eqref{eqadsmacro1} is valid for $\gamma>1/t_\eta$. Rearranging such a condition with Eqs.~\eqref{eqrate} and \eqref{eqtime1}, we obtain a transition from diffusion- to advection- dominated adsorption for:
\begin{equation}
\ell_\xi^*  \gtrsim 4 / (\gamma^* \textit{Pe}) \ . \label{eqtrans}
\end{equation}
In Fig.~\ref{fig6} we report the numerical computation of $\langle c^* \rangle_{S_\xi}$, which shows that Eq.~\eqref{eqtrans} well predicts the change from diffusion- to advection- dominated adsorption. One can interpret such a transition imagining that, for $\ell_\xi^*  \gtrsim 4 / (\gamma^* \textit{Pe})$, concentration plumes reaches the next adsorbers, at a distance $\ell_\xi$ ,  before delivering enough molecule via diffusion to establish a reaction-limited regime in the vicinity of the previous adsorber. In this condition, it is the number of adsorbers reached by the plume, which scales $\propto t$, that determines the global adsorption.

From a practical point of view, it is interesting to calculate the amount of adsorbers that yields the transition for the typical fluid-dynamic conditions characterising saturated porous susburfaces, such as soils, i.e. $\textit{Pe}=O(1)$, provided that the typical surface kinetics of the adsorbers is $\textit{Da}=O(1)$. Such a calculation, via Eq.~\eqref{eqtrans}, leads to determine the amount of particle adsorbers as roughly 5 per cubic centimetre, with higher fractions leading diffusion-dominated adsorption and lower fractions a more rapid, advection-dominated, adsorption kinetics. By equating Eqs.~\eqref{eqadsmacro1} and \eqref{eqadsmacro2}, we also conclude that a higher effective adsorption for the low-fraction configuration is experienced  at a characteristic time $t>\textit{Pe}^{-2}  t_d$, where $t_d$ is the characteristic pore-scale diffusion time $t_d=d^2/D_m$.

\section{Conclusions}

Via pore scale lattice Boltzmann simulations, we have investigated the transport and adsorption of a scalar through a chemically heterogeneous, partially adsorbing, porous medium.  
We have generated a packed bed microstructure of monodisperse spherical particles, of which a randomly chosen fraction $\xi$ is capable of adsorption. Such adsorbers are placed randomly at an average distance $\ell_\xi$ We have looked at the dynamics of a solute continuously injected into the medium and adsorbed on the fluid-solid surfaces of such adsorbing particles, focusing on the quantification of the deformation of the transported plumes of solute and limiting the analysis to fluid-dynamic conditions characterizing the transport of contaminant into subsurfaces, i.e. at $\textit{Pe}= O(1)$, and mimicking the adsorption process in a typical carbonaceous particle as biochar, i.e. setting the ratio between adsorbing and diffusive rates as $\textit{Da}=O(1)$.

We have measured the dynamical shaping of a scalar element, the solute isoscalar sheet $\Sigma$ corresponding to $c^*=1/2$, that is, half the inlet concentration value. We have followed such a dynamic process with the intent of performing a quantitative measurement of the deformation that scalar elements, such as a pocket of contaminant molecule, are subjected to, when injected into chemically heterogeneous porous subsurfaces. We have found that concentration plumes, embedded within $\Sigma$, experiences an adsorption-induced stretching process linearly dependent on time, whose rate is inversely proportional to the average distance between adsorbers, i.e. $\gamma \propto 1/\ell_\xi$. 

Following the evolution of the concentration gradients at the plume scale, we have identified two regimes: (i) a diffusion-dominated regime at early times, where the scalar elements are subjected to diffusion broadening, and (ii) an adsorption-dominated regime, where an equilibrium width $\sigma \propto \ell_\xi$ is determined. In particular, we found that the second regime onset corresponds to the characteristic diffusive time $t_\eta \propto \ell_\xi^2 $ which dissipates concentration differences in the proximity of the adsorbers. 

We have unveiled how the effects of advection-sustained plume stretching and diffusion broadening concurs to determine the macroscopic adsorption behaviour. In particular, we found that a transition between diffusion- to advection- dominated adsorption, with respective rates $\propto \sqrt{t}$ and $\propto t$, is observed by diminishing the fraction of adsorbers and we have quantified this transition at $\ell_\xi^*  \gtrsim 4 / (\gamma^* \textit{Pe})$, roughly equivalent to 5 adsorbing particles per cubic centimetre under the investigated fluid dynamic conditions. This transition should find interesting applications within a design perspective devoted to the optimisation of contaminant retention capabilities in porous subsurfaces.

We finally note that the observed transition  should hold as long as the Péclet number is from low to moderate, the medium is weakly heterogeneous from a geometrical point of view, and the characteristic adsorption kinetics is comparable with the pore-scale transport, that is $\textit{Da}=O(1)$. These observations open up rooms for further studies along these research directions, possibly with higher Péclet and Damkhöler numbers.

\appendix

\section{Lattice-Boltzmann adsorption scheme\label{app:lbm}}

We use the lattice Boltzmann method (LBM) to solve the transport and adsorption equations defined in Eqs.~\eqref{eqads} and \eqref{eqade}. The LBM is an alternative way to solve Navier-Stokes equations that exhibits sizable computational benefits when dealing with complex geometries such as porous media~\cite{succi2001lattice}. The LBM solves the local momentum transport equation by projecting the discretized Boltzmann equation along the discrete lattice directions $r$. It reads as:
\begin{eqnarray}
f_r (\mathbf{x} + \mathbf{c}_r, t+1) - f_r(\mathbf{x},t) =  - \tau_\nu^{-1} ( f_r(\mathbf{x},t)-f^{eq}_r (\mathbf{x},t)  )  + F_r \ ,
\label{eqlbm}
\end{eqnarray}
where $f_r(\mathbf{x},t)$ is the distribution function  at the position $\mathbf{x}=(x,y,z)$ and time $t$ along the $r$-th direction, $\mathbf{c}_r$ is the discrete velocity vector along the $r$-th direction, $\tau_\nu$ is the relaxation time  (proportional to fluid viscosity). With $f_r^{eq}$ we indicate the equilibrium distribution function along the $r$-th direction:
\begin{equation}
f^{eq}_r (\mathbf x,t) = w_r \rho\ \Bigg ( 1 +\frac{c_{rj} u_j(\mathbf x,t)}{c_s^2} + \frac{(c_{rj} u_j(\mathbf x,t))^2}{c_s^4} - \frac{u_j^2(\mathbf x,t)}{2c_s^2} \Bigg ) \ ,\label{eqf} 
\end{equation}
where $c_s$ is the speed of sound and $w_r$ the D3Q19 weight parameter of the three-dimensional lattice structure along the $r$-th direction.
The solution of the fluid field is deduced in each computational cell by integrating the hydrodynamic moments of the distribution functions. We can thus calculate the steady state velocity vector $u_j(\mathbf{x})$, and density $\rho(\mathbf{x})$  as:
\begin{eqnarray}
\rho(\mathbf x)  &=& \sum_r f_r(\mathbf x)   \\ 
\rho(\mathbf x)  {u}_j (\mathbf x) &=& \sum_r {c}_{rj} f_r(\mathbf x)  + \frac{1}{2}\bigg ( \frac{\Delta P}{L} \bigg )_j\ .
\end{eqnarray}
In the case of low Mach numbers, the density can be considered constant and the solution of the momentum transport equation, provided by Eq.~\ref{eqlbm}, exact with second order accuracy~\cite{succi2001lattice,guo2002discrete}.

The first step to solve Eq.~\eqref{eqade} consists of computing the steady-state solution for an incompressible fluid flowing through a given porous matrix, that is to calculate $u_j(\mathbf{x})$, via Eq.~\ref{eqlbm}. We apply a pressure gradient ${\Delta P/}{L}$ that forces the fluid through the porous microstructure, modeling it via an equivalent body force $F_r$ inserted in Eq.~\ref{eqlbm} as:
\begin{equation}
F_r (\mathbf x,t)= \Bigg ( 1-\frac{1}{2\tau_\nu} \Bigg ) w_r \Bigg ( \frac{{c}_{rj}-{u}_j(\mathbf x,t)}{c_s^2} +\frac{{c}_{rj}{u}_j(\mathbf x,t)}{c_s^4}{c}_{rj} \Bigg ) \ \bigg ( \frac{\Delta P}{L} \bigg )_r \ .
\end{equation}
Along the streamwise direction $z$ , the porous domain is extended, to straighten the flow after it exits the porous medium and avoid nonphysical effects at the $z$ border of the samples. No-slip, free-slip, and periodic boundary conditions are imposed at the fluid-solid interface, at the $x$ an $y$ transverse boundaries, and along the streamiswe direction $z$, respectively. 

As a second step, we solve a second LBM transport equation, which provide the solution of the scalar concentration field $c(\mathbf{x},t)$ transported by the underlying flow $u_j(\mathbf{x})$, see also~\cite{maggiolo2020solute}. The solute is injected at the inlet face of the samples with a step input change of concentration $c_{in}$. From the first hydrodynamic moment of this second lattice population--which we address as $g_r $--the local concentration is then extracted:
\begin{equation}
c(\mathbf x,t)= \sum_r g_r(\mathbf x,t) \ .
\end{equation}

We impose Neumann boundary condition at the adsorbing particle surfaces for the scalar lattice Boltzmann quantity $g_r$, in order  to solve Eq.~\eqref{eqads}. The distribution function at a fluid node $\mathbf{x}_a$ in the proximity of an adsorbing surface placed at  $(\mathbf{x}_a-\mathbf{c}_r)$ is corrected along the wall-normal direction $r$ as:
\begin{equation}
g_r(\mathbf{x}_a,t+1)=\dfrac{-A_1+A_2}{A_1 + A_2} g_r(\mathbf{x}_a,t) + \dfrac{2 w_r A_3}{A_1+A_2} \  , \label{eqbounce}
\end{equation}
where, for recovering the adsorption Eq.~\eqref{eqads}, the parameters are set as $A_1= k$, $A_2=1/3$ and $A_3=0$~\cite{huang2015boundary}. 
We validate the algorithm by solving the transport of a scalar quantity $c^*$ between two parallel plates, of which the bottom one adsorbs the solute with a rate $k$. In the upper boundary, zero-flux conditions are instead imposed. We considered two cases, with dimensionless adsorption rates given by $\textit{Da}=kd^2/D_m=1$ and $10$. In Fig.~\ref{fig6} we report the results, which shows the excellent agreement between numerical data and the analytical solution reported in~\citet{zhang2012general}.

\begin{figure}
\noindent\includegraphics[width=0.99\textwidth]{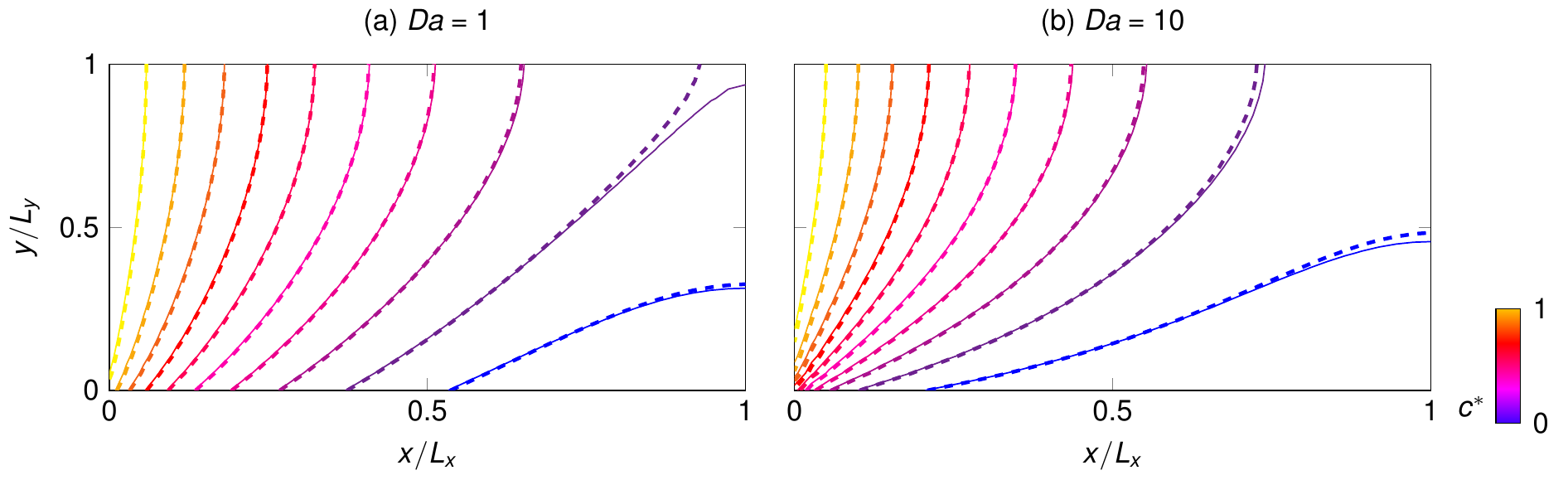}
\caption{The validation tests are set on a rectangular domain of length $L_x=40$ and height $L_y=20$ computational nodes. Two parallel plates are placed at $y=0$ and $y=L$. At the initial time, the concentration field is $c^*(x=0)=1$ at the inlet  and $c^*(x>0)=0$ elsewhere. The lower boundary adsorbs the diffusing scalar $c^*$ according to the scheme provided in Eq.~\eqref{eqbounce} and the continumm Eq.~\eqref{eqads}, for $\textit{Da}=1$, left panel (a), and $\textit{Da}=10$, right panel (b).}
\label{figS}
\end{figure}

\begin{acknowledgments}
This work was supported by the Swedish Research Council for Environment, Agricultural Sciences and Spatial Planning (FORMAS), grant number 2019–01261. The computations were enabled by resources provided by the Swedish National Infrastructure for Computing (SNIC), partially funded by the Swedish Research Council through grant agreement No. 2018–05973.
\end{acknowledgments}
%
\bibliography{apstemplate}

\begin{thebibliography}{32}%
\makeatletter
\providecommand \@ifxundefined [1]{%
 \@ifx{#1\undefined}
}%
\providecommand \@ifnum [1]{%
 \ifnum #1\expandafter \@firstoftwo
 \else \expandafter \@secondoftwo
 \fi
}%
\providecommand \@ifx [1]{%
 \ifx #1\expandafter \@firstoftwo
 \else \expandafter \@secondoftwo
 \fi
}%
\providecommand \natexlab [1]{#1}%
\providecommand \enquote  [1]{``#1''}%
\providecommand \bibnamefont  [1]{#1}%
\providecommand \bibfnamefont [1]{#1}%
\providecommand \citenamefont [1]{#1}%
\providecommand \href@noop [0]{\@secondoftwo}%
\providecommand \href [0]{\begingroup \@sanitize@url \@href}%
\providecommand \@href[1]{\@@startlink{#1}\@@href}%
\providecommand \@@href[1]{\endgroup#1\@@endlink}%
\providecommand \@sanitize@url [0]{\catcode `\\12\catcode `\$12\catcode
  `\&12\catcode `\#12\catcode `\^12\catcode `\_12\catcode `\%12\relax}%
\providecommand \@@startlink[1]{}%
\providecommand \@@endlink[0]{}%
\providecommand \url  [0]{\begingroup\@sanitize@url \@url }%
\providecommand \@url [1]{\endgroup\@href {#1}{\urlprefix }}%
\providecommand \urlprefix  [0]{URL }%
\providecommand \Eprint [0]{\href }%
\providecommand \doibase [0]{https://doi.org/}%
\providecommand \selectlanguage [0]{\@gobble}%
\providecommand \bibinfo  [0]{\@secondoftwo}%
\providecommand \bibfield  [0]{\@secondoftwo}%
\providecommand \translation [1]{[#1]}%
\providecommand \BibitemOpen [0]{}%
\providecommand \bibitemStop [0]{}%
\providecommand \bibitemNoStop [0]{.\EOS\space}%
\providecommand \EOS [0]{\spacefactor3000\relax}%
\providecommand \BibitemShut  [1]{\csname bibitem#1\endcsname}%
\let\auto@bib@innerbib\@empty
\bibitem [{\citenamefont {Berndtsson}(2010)}]{berndtsson2010green}%
  \BibitemOpen
  \bibfield  {author} {\bibinfo {author} {\bibfnamefont {J.~C.}\ \bibnamefont
  {Berndtsson}},\ }\bibfield  {title} {\bibinfo {title} {Green roof performance
  towards management of runoff water quantity and quality: A review},\
  }\href@noop {} {\bibfield  {journal} {\bibinfo  {journal} {Ecological
  engineering}\ }\textbf {\bibinfo {volume} {36}},\ \bibinfo {pages} {351}
  (\bibinfo {year} {2010})}\BibitemShut {NoStop}%
\bibitem [{\citenamefont {Kok}\ \emph {et~al.}(2016)\citenamefont {Kok},
  \citenamefont {Mohd~Sidek}, \citenamefont {Chow}, \citenamefont
  {Zainal~Abidin}, \citenamefont {Basri},\ and\ \citenamefont
  {Hayder}}]{kok2016evaluation}%
  \BibitemOpen
  \bibfield  {author} {\bibinfo {author} {\bibfnamefont {K.~H.}\ \bibnamefont
  {Kok}}, \bibinfo {author} {\bibfnamefont {L.}~\bibnamefont {Mohd~Sidek}},
  \bibinfo {author} {\bibfnamefont {M.~F.}\ \bibnamefont {Chow}}, \bibinfo
  {author} {\bibfnamefont {M.~R.}\ \bibnamefont {Zainal~Abidin}}, \bibinfo
  {author} {\bibfnamefont {H.}~\bibnamefont {Basri}},\ and\ \bibinfo {author}
  {\bibfnamefont {G.}~\bibnamefont {Hayder}},\ }\bibfield  {title} {\bibinfo
  {title} {Evaluation of green roof performances for urban stormwater quantity
  and quality controls},\ }\href@noop {} {\bibfield  {journal} {\bibinfo
  {journal} {International Journal of River Basin Management}\ }\textbf
  {\bibinfo {volume} {14}},\ \bibinfo {pages} {1} (\bibinfo {year}
  {2016})}\BibitemShut {NoStop}%
\bibitem [{\citenamefont {Ahmad}\ \emph {et~al.}(2014)\citenamefont {Ahmad},
  \citenamefont {Rajapaksha}, \citenamefont {Lim}, \citenamefont {Zhang},
  \citenamefont {Bolan}, \citenamefont {Mohan}, \citenamefont {Vithanage},
  \citenamefont {Lee},\ and\ \citenamefont {Ok}}]{ahmad2014biochar}%
  \BibitemOpen
  \bibfield  {author} {\bibinfo {author} {\bibfnamefont {M.}~\bibnamefont
  {Ahmad}}, \bibinfo {author} {\bibfnamefont {A.~U.}\ \bibnamefont
  {Rajapaksha}}, \bibinfo {author} {\bibfnamefont {J.~E.}\ \bibnamefont {Lim}},
  \bibinfo {author} {\bibfnamefont {M.}~\bibnamefont {Zhang}}, \bibinfo
  {author} {\bibfnamefont {N.}~\bibnamefont {Bolan}}, \bibinfo {author}
  {\bibfnamefont {D.}~\bibnamefont {Mohan}}, \bibinfo {author} {\bibfnamefont
  {M.}~\bibnamefont {Vithanage}}, \bibinfo {author} {\bibfnamefont {S.~S.}\
  \bibnamefont {Lee}},\ and\ \bibinfo {author} {\bibfnamefont {Y.~S.}\
  \bibnamefont {Ok}},\ }\bibfield  {title} {\bibinfo {title} {Biochar as a
  sorbent for contaminant management in soil and water: a review},\ }\href@noop
  {} {\bibfield  {journal} {\bibinfo  {journal} {Chemosphere}\ }\textbf
  {\bibinfo {volume} {99}},\ \bibinfo {pages} {19} (\bibinfo {year}
  {2014})}\BibitemShut {NoStop}%
\bibitem [{\citenamefont {Qiu}\ \emph {et~al.}(2022)\citenamefont {Qiu},
  \citenamefont {Liu}, \citenamefont {Ling}, \citenamefont {Cai}, \citenamefont
  {Yu}, \citenamefont {Wang}, \citenamefont {Fu}, \citenamefont {Hu},\ and\
  \citenamefont {Wang}}]{qiu2022biochar}%
  \BibitemOpen
  \bibfield  {author} {\bibinfo {author} {\bibfnamefont {M.}~\bibnamefont
  {Qiu}}, \bibinfo {author} {\bibfnamefont {L.}~\bibnamefont {Liu}}, \bibinfo
  {author} {\bibfnamefont {Q.}~\bibnamefont {Ling}}, \bibinfo {author}
  {\bibfnamefont {Y.}~\bibnamefont {Cai}}, \bibinfo {author} {\bibfnamefont
  {S.}~\bibnamefont {Yu}}, \bibinfo {author} {\bibfnamefont {S.}~\bibnamefont
  {Wang}}, \bibinfo {author} {\bibfnamefont {D.}~\bibnamefont {Fu}}, \bibinfo
  {author} {\bibfnamefont {B.}~\bibnamefont {Hu}},\ and\ \bibinfo {author}
  {\bibfnamefont {X.}~\bibnamefont {Wang}},\ }\bibfield  {title} {\bibinfo
  {title} {Biochar for the removal of contaminants from soil and water: A
  review},\ }\href@noop {} {\bibfield  {journal} {\bibinfo  {journal}
  {Biochar}\ }\textbf {\bibinfo {volume} {4}},\ \bibinfo {pages} {1} (\bibinfo
  {year} {2022})}\BibitemShut {NoStop}%
\bibitem [{\citenamefont {Mohanty}\ \emph {et~al.}(2018)\citenamefont
  {Mohanty}, \citenamefont {Valenca}, \citenamefont {Berger}, \citenamefont
  {Iris}, \citenamefont {Xiong}, \citenamefont {Saunders},\ and\ \citenamefont
  {Tsang}}]{mohanty2018plenty}%
  \BibitemOpen
  \bibfield  {author} {\bibinfo {author} {\bibfnamefont {S.~K.}\ \bibnamefont
  {Mohanty}}, \bibinfo {author} {\bibfnamefont {R.}~\bibnamefont {Valenca}},
  \bibinfo {author} {\bibfnamefont {A.~W.}\ \bibnamefont {Berger}}, \bibinfo
  {author} {\bibfnamefont {K.}~\bibnamefont {Iris}}, \bibinfo {author}
  {\bibfnamefont {X.}~\bibnamefont {Xiong}}, \bibinfo {author} {\bibfnamefont
  {T.~M.}\ \bibnamefont {Saunders}},\ and\ \bibinfo {author} {\bibfnamefont
  {D.~C.}\ \bibnamefont {Tsang}},\ }\bibfield  {title} {\bibinfo {title}
  {Plenty of room for carbon on the ground: potential applications of biochar
  for stormwater treatment},\ }\href@noop {} {\bibfield  {journal} {\bibinfo
  {journal} {Science of the total environment}\ }\textbf {\bibinfo {volume}
  {625}},\ \bibinfo {pages} {1644} (\bibinfo {year} {2018})}\BibitemShut
  {NoStop}%
\bibitem [{\citenamefont {Villermaux}(2012)}]{villermaux2012mixing}%
  \BibitemOpen
  \bibfield  {author} {\bibinfo {author} {\bibfnamefont {E.}~\bibnamefont
  {Villermaux}},\ }\bibfield  {title} {\bibinfo {title} {Mixing by porous
  media},\ }\href@noop {} {\bibfield  {journal} {\bibinfo  {journal} {Comptes
  Rendus M{\'e}canique}\ }\textbf {\bibinfo {volume} {340}},\ \bibinfo {pages}
  {933} (\bibinfo {year} {2012})}\BibitemShut {NoStop}%
\bibitem [{\citenamefont {Le~Borgne}\ \emph {et~al.}(2015)\citenamefont
  {Le~Borgne}, \citenamefont {Dentz},\ and\ \citenamefont
  {Villermaux}}]{le2015lamellar}%
  \BibitemOpen
  \bibfield  {author} {\bibinfo {author} {\bibfnamefont {T.}~\bibnamefont
  {Le~Borgne}}, \bibinfo {author} {\bibfnamefont {M.}~\bibnamefont {Dentz}},\
  and\ \bibinfo {author} {\bibfnamefont {E.}~\bibnamefont {Villermaux}},\
  }\bibfield  {title} {\bibinfo {title} {The lamellar description of mixing in
  porous media},\ }\href@noop {} {\bibfield  {journal} {\bibinfo  {journal}
  {Journal of Fluid Mechanics}\ }\textbf {\bibinfo {volume} {770}},\ \bibinfo
  {pages} {458} (\bibinfo {year} {2015})}\BibitemShut {NoStop}%
\bibitem [{\citenamefont {Lester}\ \emph {et~al.}(2013)\citenamefont {Lester},
  \citenamefont {Metcalfe},\ and\ \citenamefont {Trefry}}]{lester2013chaotic}%
  \BibitemOpen
  \bibfield  {author} {\bibinfo {author} {\bibfnamefont {D.~R.}\ \bibnamefont
  {Lester}}, \bibinfo {author} {\bibfnamefont {G.}~\bibnamefont {Metcalfe}},\
  and\ \bibinfo {author} {\bibfnamefont {M.~G.}\ \bibnamefont {Trefry}},\
  }\bibfield  {title} {\bibinfo {title} {Is chaotic advection inherent to
  porous media flow?},\ }\href {https://doi.org/10.1103/PhysRevLett.111.174101}
  {\bibfield  {journal} {\bibinfo  {journal} {Phys. Rev. Lett.}\ }\textbf
  {\bibinfo {volume} {111}},\ \bibinfo {pages} {174101} (\bibinfo {year}
  {2013})}\BibitemShut {NoStop}%
\bibitem [{\citenamefont {Aref}\ \emph {et~al.}(2017)\citenamefont {Aref},
  \citenamefont {Blake}, \citenamefont {Budi\ifmmode \check{s}\else
  \v{s}\fi{}i\ifmmode~\acute{c}\else \'{c}\fi{}}, \citenamefont {Cardoso},
  \citenamefont {Cartwright}, \citenamefont {Clercx}, \citenamefont {El~Omari},
  \citenamefont {Feudel}, \citenamefont {Golestanian}, \citenamefont
  {Gouillart}, \citenamefont {van Heijst}, \citenamefont {Krasnopolskaya},
  \citenamefont {Le~Guer}, \citenamefont {MacKay}, \citenamefont {Meleshko},
  \citenamefont {Metcalfe}, \citenamefont {Mezi\ifmmode~\acute{c}\else
  \'{c}\fi{}}, \citenamefont {de~Moura}, \citenamefont {Piro}, \citenamefont
  {Speetjens}, \citenamefont {Sturman}, \citenamefont {Thiffeault},\ and\
  \citenamefont {Tuval}}]{aref2017frontiers}%
  \BibitemOpen
  \bibfield  {author} {\bibinfo {author} {\bibfnamefont {H.}~\bibnamefont
  {Aref}}, \bibinfo {author} {\bibfnamefont {J.~R.}\ \bibnamefont {Blake}},
  \bibinfo {author} {\bibfnamefont {M.}~\bibnamefont {Budi\ifmmode
  \check{s}\else \v{s}\fi{}i\ifmmode~\acute{c}\else \'{c}\fi{}}}, \bibinfo
  {author} {\bibfnamefont {S.~S.~S.}\ \bibnamefont {Cardoso}}, \bibinfo
  {author} {\bibfnamefont {J.~H.~E.}\ \bibnamefont {Cartwright}}, \bibinfo
  {author} {\bibfnamefont {H.~J.~H.}\ \bibnamefont {Clercx}}, \bibinfo {author}
  {\bibfnamefont {K.}~\bibnamefont {El~Omari}}, \bibinfo {author}
  {\bibfnamefont {U.}~\bibnamefont {Feudel}}, \bibinfo {author} {\bibfnamefont
  {R.}~\bibnamefont {Golestanian}}, \bibinfo {author} {\bibfnamefont
  {E.}~\bibnamefont {Gouillart}}, \bibinfo {author} {\bibfnamefont {G.~F.}\
  \bibnamefont {van Heijst}}, \bibinfo {author} {\bibfnamefont {T.~S.}\
  \bibnamefont {Krasnopolskaya}}, \bibinfo {author} {\bibfnamefont
  {Y.}~\bibnamefont {Le~Guer}}, \bibinfo {author} {\bibfnamefont {R.~S.}\
  \bibnamefont {MacKay}}, \bibinfo {author} {\bibfnamefont {V.~V.}\
  \bibnamefont {Meleshko}}, \bibinfo {author} {\bibfnamefont {G.}~\bibnamefont
  {Metcalfe}}, \bibinfo {author} {\bibfnamefont {I.}~\bibnamefont
  {Mezi\ifmmode~\acute{c}\else \'{c}\fi{}}}, \bibinfo {author} {\bibfnamefont
  {A.~P.~S.}\ \bibnamefont {de~Moura}}, \bibinfo {author} {\bibfnamefont
  {O.}~\bibnamefont {Piro}}, \bibinfo {author} {\bibfnamefont {M.~F.~M.}\
  \bibnamefont {Speetjens}}, \bibinfo {author} {\bibfnamefont {R.}~\bibnamefont
  {Sturman}}, \bibinfo {author} {\bibfnamefont {J.-L.}\ \bibnamefont
  {Thiffeault}},\ and\ \bibinfo {author} {\bibfnamefont {I.}~\bibnamefont
  {Tuval}},\ }\bibfield  {title} {\bibinfo {title} {Frontiers of chaotic
  advection},\ }\href {https://doi.org/10.1103/RevModPhys.89.025007} {\bibfield
   {journal} {\bibinfo  {journal} {Rev. Mod. Phys.}\ }\textbf {\bibinfo
  {volume} {89}},\ \bibinfo {pages} {025007} (\bibinfo {year}
  {2017})}\BibitemShut {NoStop}%
\bibitem [{\citenamefont {Heyman}\ \emph {et~al.}(2020)\citenamefont {Heyman},
  \citenamefont {Lester}, \citenamefont {Turuban}, \citenamefont
  {M{\'e}heust},\ and\ \citenamefont {Le~Borgne}}]{heyman2020stretching}%
  \BibitemOpen
  \bibfield  {author} {\bibinfo {author} {\bibfnamefont {J.}~\bibnamefont
  {Heyman}}, \bibinfo {author} {\bibfnamefont {D.~R.}\ \bibnamefont {Lester}},
  \bibinfo {author} {\bibfnamefont {R.}~\bibnamefont {Turuban}}, \bibinfo
  {author} {\bibfnamefont {Y.}~\bibnamefont {M{\'e}heust}},\ and\ \bibinfo
  {author} {\bibfnamefont {T.}~\bibnamefont {Le~Borgne}},\ }\bibfield  {title}
  {\bibinfo {title} {Stretching and folding sustain microscale chemical
  gradients in porous media},\ }\href@noop {} {\bibfield  {journal} {\bibinfo
  {journal} {Proceedings of the National Academy of Sciences}\ }\textbf
  {\bibinfo {volume} {117}},\ \bibinfo {pages} {13359} (\bibinfo {year}
  {2020})}\BibitemShut {NoStop}%
\bibitem [{\citenamefont {Liu}\ \emph {et~al.}(2019)\citenamefont {Liu},
  \citenamefont {Feng}, \citenamefont {Chen}, \citenamefont {Wei},\ and\
  \citenamefont {Deo}}]{liu2019influence}%
  \BibitemOpen
  \bibfield  {author} {\bibinfo {author} {\bibfnamefont {W.}~\bibnamefont
  {Liu}}, \bibinfo {author} {\bibfnamefont {Q.}~\bibnamefont {Feng}}, \bibinfo
  {author} {\bibfnamefont {W.}~\bibnamefont {Chen}}, \bibinfo {author}
  {\bibfnamefont {W.}~\bibnamefont {Wei}},\ and\ \bibinfo {author}
  {\bibfnamefont {R.~C.}\ \bibnamefont {Deo}},\ }\bibfield  {title} {\bibinfo
  {title} {The influence of structural factors on stormwater runoff retention
  of extensive green roofs: new evidence from scale-based models and real
  experiments},\ }\href@noop {} {\bibfield  {journal} {\bibinfo  {journal}
  {Journal of Hydrology}\ }\textbf {\bibinfo {volume} {569}},\ \bibinfo {pages}
  {230} (\bibinfo {year} {2019})}\BibitemShut {NoStop}%
\bibitem [{\citenamefont {Olsson}\ \emph {et~al.}(2018)\citenamefont {Olsson},
  \citenamefont {Berg}, \citenamefont {Eronn}, \citenamefont {Simonsson},
  \citenamefont {S{\"o}dling}, \citenamefont {Wern},\ and\ \citenamefont
  {Yang}}]{olsson2018extremregn}%
  \BibitemOpen
  \bibfield  {author} {\bibinfo {author} {\bibfnamefont {J.}~\bibnamefont
  {Olsson}}, \bibinfo {author} {\bibfnamefont {P.}~\bibnamefont {Berg}},
  \bibinfo {author} {\bibfnamefont {A.}~\bibnamefont {Eronn}}, \bibinfo
  {author} {\bibfnamefont {L.}~\bibnamefont {Simonsson}}, \bibinfo {author}
  {\bibfnamefont {J.}~\bibnamefont {S{\"o}dling}}, \bibinfo {author}
  {\bibfnamefont {L.}~\bibnamefont {Wern}},\ and\ \bibinfo {author}
  {\bibfnamefont {W.}~\bibnamefont {Yang}},\ }\href@noop {} {\emph {\bibinfo
  {title} {Extremregn i nuvarande och framtida klimat Analyser av observationer
  och framtidsscenarier}}}\ (\bibinfo {year} {2018})\BibitemShut {NoStop}%
\bibitem [{\citenamefont {Heyman}\ \emph {et~al.}(2021)\citenamefont {Heyman},
  \citenamefont {Lester},\ and\ \citenamefont {Le~Borgne}}]{heyman2021scalar}%
  \BibitemOpen
  \bibfield  {author} {\bibinfo {author} {\bibfnamefont {J.}~\bibnamefont
  {Heyman}}, \bibinfo {author} {\bibfnamefont {D.~R.}\ \bibnamefont {Lester}},\
  and\ \bibinfo {author} {\bibfnamefont {T.}~\bibnamefont {Le~Borgne}},\
  }\bibfield  {title} {\bibinfo {title} {Scalar signatures of chaotic mixing in
  porous media},\ }\href {https://doi.org/10.1103/PhysRevLett.126.034505}
  {\bibfield  {journal} {\bibinfo  {journal} {Phys. Rev. Lett.}\ }\textbf
  {\bibinfo {volume} {126}},\ \bibinfo {pages} {034505} (\bibinfo {year}
  {2021})}\BibitemShut {NoStop}%
\bibitem [{\citenamefont {Kree}\ and\ \citenamefont
  {Villermaux}(2017)}]{kree2017scalar}%
  \BibitemOpen
  \bibfield  {author} {\bibinfo {author} {\bibfnamefont {M.}~\bibnamefont
  {Kree}}\ and\ \bibinfo {author} {\bibfnamefont {E.}~\bibnamefont
  {Villermaux}},\ }\bibfield  {title} {\bibinfo {title} {Scalar mixtures in
  porous media},\ }\href@noop {} {\bibfield  {journal} {\bibinfo  {journal}
  {Physical Review Fluids}\ }\textbf {\bibinfo {volume} {2}},\ \bibinfo {pages}
  {104502} (\bibinfo {year} {2017})}\BibitemShut {NoStop}%
\bibitem [{\citenamefont {Turuban}\ \emph {et~al.}(2019)\citenamefont
  {Turuban}, \citenamefont {Lester}, \citenamefont {Heyman}, \citenamefont
  {Le~Borgne},\ and\ \citenamefont {M{\'e}heust}}]{turuban2019chaotic}%
  \BibitemOpen
  \bibfield  {author} {\bibinfo {author} {\bibfnamefont {R.}~\bibnamefont
  {Turuban}}, \bibinfo {author} {\bibfnamefont {D.~R.}\ \bibnamefont {Lester}},
  \bibinfo {author} {\bibfnamefont {J.}~\bibnamefont {Heyman}}, \bibinfo
  {author} {\bibfnamefont {T.}~\bibnamefont {Le~Borgne}},\ and\ \bibinfo
  {author} {\bibfnamefont {Y.}~\bibnamefont {M{\'e}heust}},\ }\bibfield
  {title} {\bibinfo {title} {Chaotic mixing in crystalline granular media},\
  }\href@noop {} {\bibfield  {journal} {\bibinfo  {journal} {Journal of Fluid
  Mechanics}\ }\textbf {\bibinfo {volume} {871}},\ \bibinfo {pages} {562}
  (\bibinfo {year} {2019})}\BibitemShut {NoStop}%
\bibitem [{\citenamefont {Souzy}\ \emph {et~al.}(2020)\citenamefont {Souzy},
  \citenamefont {Lhuissier}, \citenamefont {M{\'e}heust}, \citenamefont
  {Le~Borgne},\ and\ \citenamefont {Metzger}}]{souzy2020velocity}%
  \BibitemOpen
  \bibfield  {author} {\bibinfo {author} {\bibfnamefont {M.}~\bibnamefont
  {Souzy}}, \bibinfo {author} {\bibfnamefont {H.}~\bibnamefont {Lhuissier}},
  \bibinfo {author} {\bibfnamefont {Y.}~\bibnamefont {M{\'e}heust}}, \bibinfo
  {author} {\bibfnamefont {T.}~\bibnamefont {Le~Borgne}},\ and\ \bibinfo
  {author} {\bibfnamefont {B.}~\bibnamefont {Metzger}},\ }\bibfield  {title}
  {\bibinfo {title} {Velocity distributions, dispersion and stretching in
  three-dimensional porous media},\ }\href@noop {} {\bibfield  {journal}
  {\bibinfo  {journal} {Journal of Fluid Mechanics}\ }\textbf {\bibinfo
  {volume} {891}} (\bibinfo {year} {2020})}\BibitemShut {NoStop}%
\bibitem [{\citenamefont {Ye}\ \emph {et~al.}(2020)\citenamefont {Ye},
  \citenamefont {Chiogna}, \citenamefont {Lu},\ and\ \citenamefont
  {Rolle}}]{ye2020plume}%
  \BibitemOpen
  \bibfield  {author} {\bibinfo {author} {\bibfnamefont {Y.}~\bibnamefont
  {Ye}}, \bibinfo {author} {\bibfnamefont {G.}~\bibnamefont {Chiogna}},
  \bibinfo {author} {\bibfnamefont {C.}~\bibnamefont {Lu}},\ and\ \bibinfo
  {author} {\bibfnamefont {M.}~\bibnamefont {Rolle}},\ }\bibfield  {title}
  {\bibinfo {title} {Plume deformation, mixing, and reaction kinetics: An
  analysis of interacting helical flows in three-dimensional porous media},\
  }\href@noop {} {\bibfield  {journal} {\bibinfo  {journal} {Physical Review
  E}\ }\textbf {\bibinfo {volume} {102}},\ \bibinfo {pages} {013110} (\bibinfo
  {year} {2020})}\BibitemShut {NoStop}%
\bibitem [{\citenamefont {Miele}\ \emph {et~al.}(2019)\citenamefont {Miele},
  \citenamefont {de~Anna},\ and\ \citenamefont {Dentz}}]{miele2019stochastic}%
  \BibitemOpen
  \bibfield  {author} {\bibinfo {author} {\bibfnamefont {F.}~\bibnamefont
  {Miele}}, \bibinfo {author} {\bibfnamefont {P.}~\bibnamefont {de~Anna}},\
  and\ \bibinfo {author} {\bibfnamefont {M.}~\bibnamefont {Dentz}},\ }\bibfield
   {title} {\bibinfo {title} {Stochastic model for filtration by porous
  materials},\ }\href {https://doi.org/10.1103/PhysRevFluids.4.094101}
  {\bibfield  {journal} {\bibinfo  {journal} {Phys. Rev. Fluids}\ }\textbf
  {\bibinfo {volume} {4}},\ \bibinfo {pages} {094101} (\bibinfo {year}
  {2019})}\BibitemShut {NoStop}%
\bibitem [{\citenamefont {Johnson}\ \emph {et~al.}(2003)\citenamefont
  {Johnson}, \citenamefont {Gupta}, \citenamefont {Putz}, \citenamefont {Hu},\
  and\ \citenamefont {Brusseau}}]{johnson2003effect}%
  \BibitemOpen
  \bibfield  {author} {\bibinfo {author} {\bibfnamefont {G.}~\bibnamefont
  {Johnson}}, \bibinfo {author} {\bibfnamefont {K.}~\bibnamefont {Gupta}},
  \bibinfo {author} {\bibfnamefont {D.}~\bibnamefont {Putz}}, \bibinfo {author}
  {\bibfnamefont {Q.}~\bibnamefont {Hu}},\ and\ \bibinfo {author}
  {\bibfnamefont {M.}~\bibnamefont {Brusseau}},\ }\bibfield  {title} {\bibinfo
  {title} {The effect of local-scale physical heterogeneity and nonlinear,
  rate-limited sorption/desorption on contaminant transport in porous media},\
  }\href@noop {} {\bibfield  {journal} {\bibinfo  {journal} {Journal of
  contaminant hydrology}\ }\textbf {\bibinfo {volume} {64}},\ \bibinfo {pages}
  {35} (\bibinfo {year} {2003})}\BibitemShut {NoStop}%
\bibitem [{\citenamefont {Tyukhova}\ \emph {et~al.}(2016)\citenamefont
  {Tyukhova}, \citenamefont {Dentz}, \citenamefont {Kinzelbach},\ and\
  \citenamefont {Willmann}}]{tyukhova2016mechanisms}%
  \BibitemOpen
  \bibfield  {author} {\bibinfo {author} {\bibfnamefont {A.}~\bibnamefont
  {Tyukhova}}, \bibinfo {author} {\bibfnamefont {M.}~\bibnamefont {Dentz}},
  \bibinfo {author} {\bibfnamefont {W.}~\bibnamefont {Kinzelbach}},\ and\
  \bibinfo {author} {\bibfnamefont {M.}~\bibnamefont {Willmann}},\ }\bibfield
  {title} {\bibinfo {title} {Mechanisms of anomalous dispersion in flow through
  heterogeneous porous media},\ }\href@noop {} {\bibfield  {journal} {\bibinfo
  {journal} {Physical Review Fluids}\ }\textbf {\bibinfo {volume} {1}},\
  \bibinfo {pages} {074002} (\bibinfo {year} {2016})}\BibitemShut {NoStop}%
\bibitem [{\citenamefont {Municchi}\ and\ \citenamefont
  {Icardi}(2020)}]{municchi2020macroscopic}%
  \BibitemOpen
  \bibfield  {author} {\bibinfo {author} {\bibfnamefont {F.}~\bibnamefont
  {Municchi}}\ and\ \bibinfo {author} {\bibfnamefont {M.}~\bibnamefont
  {Icardi}},\ }\bibfield  {title} {\bibinfo {title} {Macroscopic models for
  filtration and heterogeneous reactions in porous media},\ }\href@noop {}
  {\bibfield  {journal} {\bibinfo  {journal} {Advances in Water Resources}\
  }\textbf {\bibinfo {volume} {141}},\ \bibinfo {pages} {103605} (\bibinfo
  {year} {2020})}\BibitemShut {NoStop}%
\bibitem [{\citenamefont {Boccardo}\ \emph {et~al.}(2015)\citenamefont
  {Boccardo}, \citenamefont {Augier}, \citenamefont {Haroun}, \citenamefont
  {Ferr{\'e}},\ and\ \citenamefont {Marchisio}}]{boccardo2015validation}%
  \BibitemOpen
  \bibfield  {author} {\bibinfo {author} {\bibfnamefont {G.}~\bibnamefont
  {Boccardo}}, \bibinfo {author} {\bibfnamefont {F.}~\bibnamefont {Augier}},
  \bibinfo {author} {\bibfnamefont {Y.}~\bibnamefont {Haroun}}, \bibinfo
  {author} {\bibfnamefont {D.}~\bibnamefont {Ferr{\'e}}},\ and\ \bibinfo
  {author} {\bibfnamefont {D.~L.}\ \bibnamefont {Marchisio}},\ }\bibfield
  {title} {\bibinfo {title} {Validation of a novel open-source work-flow for
  the simulation of packed-bed reactors},\ }\href@noop {} {\bibfield  {journal}
  {\bibinfo  {journal} {Chemical Engineering Journal}\ }\textbf {\bibinfo
  {volume} {279}},\ \bibinfo {pages} {809} (\bibinfo {year}
  {2015})}\BibitemShut {NoStop}%
\bibitem [{\citenamefont {Pettersson}\ \emph {et~al.}(2020)\citenamefont
  {Pettersson}, \citenamefont {Maggiolo}, \citenamefont {Sasic}, \citenamefont
  {Johansson},\ and\ \citenamefont {Sasic-Kalagasidis}}]{pettersson2020impact}%
  \BibitemOpen
  \bibfield  {author} {\bibinfo {author} {\bibfnamefont {K.}~\bibnamefont
  {Pettersson}}, \bibinfo {author} {\bibfnamefont {D.}~\bibnamefont
  {Maggiolo}}, \bibinfo {author} {\bibfnamefont {S.}~\bibnamefont {Sasic}},
  \bibinfo {author} {\bibfnamefont {P.}~\bibnamefont {Johansson}},\ and\
  \bibinfo {author} {\bibfnamefont {A.}~\bibnamefont {Sasic-Kalagasidis}},\
  }\bibfield  {title} {\bibinfo {title} {On the impact of porous media
  microstructure on rainfall infiltration of thin homogeneous green roof growth
  substrates},\ }\href@noop {} {\bibfield  {journal} {\bibinfo  {journal}
  {Journal of Hydrology}\ }\textbf {\bibinfo {volume} {582}},\ \bibinfo {pages}
  {124286} (\bibinfo {year} {2020})}\BibitemShut {NoStop}%
\bibitem [{\citenamefont {Succi}(2001)}]{succi2001lattice}%
  \BibitemOpen
  \bibfield  {author} {\bibinfo {author} {\bibfnamefont {S.}~\bibnamefont
  {Succi}},\ }\href@noop {} {\emph {\bibinfo {title} {The lattice Boltzmann
  equation: for fluid dynamics and beyond}}}\ (\bibinfo  {publisher} {Oxford
  university press},\ \bibinfo {year} {2001})\BibitemShut {NoStop}%
\bibitem [{\citenamefont {Maggiolo}\ \emph {et~al.}(2020)\citenamefont
  {Maggiolo}, \citenamefont {Picano}, \citenamefont {Zanini}, \citenamefont
  {Carmignato}, \citenamefont {Guarnieri}, \citenamefont {Sasic},\ and\
  \citenamefont {Str{\"o}m}}]{maggiolo2020solute}%
  \BibitemOpen
  \bibfield  {author} {\bibinfo {author} {\bibfnamefont {D.}~\bibnamefont
  {Maggiolo}}, \bibinfo {author} {\bibfnamefont {F.}~\bibnamefont {Picano}},
  \bibinfo {author} {\bibfnamefont {F.}~\bibnamefont {Zanini}}, \bibinfo
  {author} {\bibfnamefont {S.}~\bibnamefont {Carmignato}}, \bibinfo {author}
  {\bibfnamefont {M.}~\bibnamefont {Guarnieri}}, \bibinfo {author}
  {\bibfnamefont {S.}~\bibnamefont {Sasic}},\ and\ \bibinfo {author}
  {\bibfnamefont {H.}~\bibnamefont {Str{\"o}m}},\ }\bibfield  {title} {\bibinfo
  {title} {Solute transport and reaction in porous electrodes at high schmidt
  numbers},\ }\href@noop {} {\bibfield  {journal} {\bibinfo  {journal} {Journal
  of Fluid Mechanics}\ }\textbf {\bibinfo {volume} {896}} (\bibinfo {year}
  {2020})}\BibitemShut {NoStop}%
\bibitem [{\citenamefont {Cermelli}\ \emph {et~al.}(2005)\citenamefont
  {Cermelli}, \citenamefont {Fried},\ and\ \citenamefont
  {Gurtin}}]{cermelli2005transport}%
  \BibitemOpen
  \bibfield  {author} {\bibinfo {author} {\bibfnamefont {P.}~\bibnamefont
  {Cermelli}}, \bibinfo {author} {\bibfnamefont {E.}~\bibnamefont {Fried}},\
  and\ \bibinfo {author} {\bibfnamefont {M.~E.}\ \bibnamefont {Gurtin}},\
  }\bibfield  {title} {\bibinfo {title} {Transport relations for surface
  integrals arising in the formulation of balance laws for evolving fluid
  interfaces},\ }\href@noop {} {\bibfield  {journal} {\bibinfo  {journal}
  {Journal of Fluid Mechanics}\ }\textbf {\bibinfo {volume} {544}},\ \bibinfo
  {pages} {339} (\bibinfo {year} {2005})}\BibitemShut {NoStop}%
\bibitem [{\citenamefont {Tufenkji}\ \emph {et~al.}(2003)\citenamefont
  {Tufenkji}, \citenamefont {Redman},\ and\ \citenamefont
  {Elimelech}}]{tufenkji2003interpreting}%
  \BibitemOpen
  \bibfield  {author} {\bibinfo {author} {\bibfnamefont {N.}~\bibnamefont
  {Tufenkji}}, \bibinfo {author} {\bibfnamefont {J.~A.}\ \bibnamefont
  {Redman}},\ and\ \bibinfo {author} {\bibfnamefont {M.}~\bibnamefont
  {Elimelech}},\ }\bibfield  {title} {\bibinfo {title} {Interpreting deposition
  patterns of microbial particles in laboratory-scale column experiments},\
  }\href@noop {} {\bibfield  {journal} {\bibinfo  {journal} {Environmental
  Science \& Technology}\ }\textbf {\bibinfo {volume} {37}},\ \bibinfo {pages}
  {616} (\bibinfo {year} {2003})}\BibitemShut {NoStop}%
\bibitem [{\citenamefont {Brenner}(2013)}]{brenner2013macrotransport}%
  \BibitemOpen
  \bibfield  {author} {\bibinfo {author} {\bibfnamefont {H.}~\bibnamefont
  {Brenner}},\ }\href@noop {} {\emph {\bibinfo {title} {Macrotransport
  processes}}}\ (\bibinfo  {publisher} {Elsevier},\ \bibinfo {year}
  {2013})\BibitemShut {NoStop}%
\bibitem [{\citenamefont {Yao}\ \emph {et~al.}(1971)\citenamefont {Yao},
  \citenamefont {Habibian},\ and\ \citenamefont {O'Melia}}]{yao1971water}%
  \BibitemOpen
  \bibfield  {author} {\bibinfo {author} {\bibfnamefont {K.-M.}\ \bibnamefont
  {Yao}}, \bibinfo {author} {\bibfnamefont {M.~T.}\ \bibnamefont {Habibian}},\
  and\ \bibinfo {author} {\bibfnamefont {C.~R.}\ \bibnamefont {O'Melia}},\
  }\bibfield  {title} {\bibinfo {title} {Water and waste water filtration.
  concepts and applications},\ }\href@noop {} {\bibfield  {journal} {\bibinfo
  {journal} {Environmental science \& technology}\ }\textbf {\bibinfo {volume}
  {5}},\ \bibinfo {pages} {1105} (\bibinfo {year} {1971})}\BibitemShut
  {NoStop}%
\bibitem [{\citenamefont {Guo}\ \emph {et~al.}(2002)\citenamefont {Guo},
  \citenamefont {Zheng},\ and\ \citenamefont {Shi}}]{guo2002discrete}%
  \BibitemOpen
  \bibfield  {author} {\bibinfo {author} {\bibfnamefont {Z.}~\bibnamefont
  {Guo}}, \bibinfo {author} {\bibfnamefont {C.}~\bibnamefont {Zheng}},\ and\
  \bibinfo {author} {\bibfnamefont {B.}~\bibnamefont {Shi}},\ }\bibfield
  {title} {\bibinfo {title} {Discrete lattice effects on the forcing term in
  the lattice boltzmann method},\ }\href@noop {} {\bibfield  {journal}
  {\bibinfo  {journal} {Physical review E}\ }\textbf {\bibinfo {volume} {65}},\
  \bibinfo {pages} {046308} (\bibinfo {year} {2002})}\BibitemShut {NoStop}%
\bibitem [{\citenamefont {Huang}\ and\ \citenamefont
  {Yong}(2015)}]{huang2015boundary}%
  \BibitemOpen
  \bibfield  {author} {\bibinfo {author} {\bibfnamefont {J.}~\bibnamefont
  {Huang}}\ and\ \bibinfo {author} {\bibfnamefont {W.-A.}\ \bibnamefont
  {Yong}},\ }\bibfield  {title} {\bibinfo {title} {Boundary conditions of the
  lattice boltzmann method for convection--diffusion equations},\ }\href@noop
  {} {\bibfield  {journal} {\bibinfo  {journal} {Journal of Computational
  Physics}\ }\textbf {\bibinfo {volume} {300}},\ \bibinfo {pages} {70}
  (\bibinfo {year} {2015})}\BibitemShut {NoStop}%
\bibitem [{\citenamefont {Zhang}\ \emph {et~al.}(2012)\citenamefont {Zhang},
  \citenamefont {Shi}, \citenamefont {Guo}, \citenamefont {Chai},\ and\
  \citenamefont {Lu}}]{zhang2012general}%
  \BibitemOpen
  \bibfield  {author} {\bibinfo {author} {\bibfnamefont {T.}~\bibnamefont
  {Zhang}}, \bibinfo {author} {\bibfnamefont {B.}~\bibnamefont {Shi}}, \bibinfo
  {author} {\bibfnamefont {Z.}~\bibnamefont {Guo}}, \bibinfo {author}
  {\bibfnamefont {Z.}~\bibnamefont {Chai}},\ and\ \bibinfo {author}
  {\bibfnamefont {J.}~\bibnamefont {Lu}},\ }\bibfield  {title} {\bibinfo
  {title} {General bounce-back scheme for concentration boundary condition in
  the lattice-boltzmann method},\ }\href@noop {} {\bibfield  {journal}
  {\bibinfo  {journal} {Physical Review E}\ }\textbf {\bibinfo {volume} {85}},\
  \bibinfo {pages} {016701} (\bibinfo {year} {2012})}\BibitemShut {NoStop}%
\end{thebibliography}%
\end{document}